\documentclass[aps,prl,twocolumn,superscriptaddress,showpacs,amsmath,amssymb]{revtex4-1}

\usepackage{graphicx}

\newcommand{\Fref}[1]{Figure~\ref{#1}}
\newcommand{\fref}[1]{Fig.~\ref{#1}}

\usepackage{amstext}

\begin{document}

\title{Effects of electron irradiation on resistivity and London penetration \\ depth of Ba$_{1-x}$K$_x$Fe$_2$As$_2$ ($x \leq$ 0.34) iron - pnictide superconductor}

\author{K.~Cho}
\affiliation{The Ames Laboratory and Department of Physics $\&$ Astronomy, Iowa State University, Ames, IA 50011, USA}

\author{M.~Ko\'nczykowski}
\affiliation{Laboratoire des Solides Irradi\'es, CNRS UMR 7642 $\&$ CEA-DSM-IRAMIS, Ecole Polytechnique, F-91128 Palaiseau cedex, France}

\author{J.~Murphy}
\affiliation{The Ames Laboratory and Department of Physics $\&$ Astronomy, Iowa State University, Ames, IA 50011, USA}

\author{H.~Kim}
\affiliation{The Ames Laboratory and Department of Physics $\&$ Astronomy, Iowa State University, Ames, IA 50011, USA}

\author{M.~A.~Tanatar}
\affiliation{The Ames Laboratory and Department of Physics $\&$ Astronomy, Iowa State University, Ames, IA 50011, USA}

\author{W.~E.~Straszheim}
\affiliation{The Ames Laboratory and Department of Physics $\&$ Astronomy, Iowa State University, Ames, IA 50011, USA}

\author{B.~Shen}
\affiliation{Center for Superconducting Physics and Materials,
National Laboratory of Solid State Microstructures $\&$ Department of Physics, Nanjing University, Nanjing 210093, China}

\author{H.~H.~Wen}
\affiliation{Center for Superconducting Physics and Materials,
National Laboratory of Solid State Microstructures $\&$ Department of Physics, Nanjing University, Nanjing 210093, China}

\author{R.~Prozorov}
\email[Corresponding author: ]{prozorov@ameslab.gov}
\affiliation{The Ames Laboratory and Department of Physics $\&$ Astronomy, Iowa State University, Ames, IA 50011, USA}

\date{28 July 2014}

\begin{abstract}
Irradiation with 2.5 MeV electrons at doses up to 5.2 $\times$10$^{19}$ electrons/cm$^2$  was used to introduce point-like defects in single crystals of Ba$_{1-x}$K$_x$Fe$_2$As$_2$ with $x=$ 0.19 ($T_c=$ 14 K), $x=$ 0.26 ($T_c=$ 32 K) and 0.34 ($T_c=$ 39 K) to study the superconducting gap structure by probing the effect of non-magnetic scattering on electrical resistivity, $\rho(T)$, and London penetration depth, $ \lambda (T)$. For all compositions, the irradiation suppressed the superconducting transition temperature, $T_c$ and increased resistivity. The low - temperature behavior of $\lambda (T)$ is best described by the power - law function, $\Delta \lambda (T) = A(T/T_c)^n$. While substantial suppression of $T_c$ supports $s_{\pm}$ pairing mechanism, in samples close to the optimal doping, $x=$ 0.26 and 0.34, the exponent $n$ remained high ($n \geq 3$) indicating robust full superconducting gaps. For the $x=$ 0.19 composition, exhibiting coexistence of superconductivity and long - range magnetism, the suppression of $T_c$ was much more rapid and the exponent $n$ decreased toward dirty limit of  $n$ = 2. In this sample, the irradiation also suppressed the temperature of structural/magnetic transition, $T_{sm}$, from 103 K to 98 K consistent with the itinerant nature of the magnetic order. Our results suggest that underdoped compositions, especially in the coexisting regime are most susceptible to non-magnetic scattering and imply that in multi-band Ba$_{1-x}$K$_x$Fe$_2$As$_2$ superconductors, the ratio of the inter-band to intra-band pairing strength, and associated gap anisotropy, increases upon the departure from the optimal doping.
\end{abstract}



\maketitle

\section{Introduction}

Studying the effects of a controlled point - like disorder on superconducting properties is a powerful tool to understand the mechanisms of superconductivity  \cite{Anderson1959JPCS,Abrikosov1960,Hirschfeld1993PRB,KimMuzikar1994PRB,Balatsky2006RMP,WangHirschfeldMishra2013PRB,Prozorov2014BaRu,Mizukami2014eprint}. According to the Anderson's theorem, conventional isotropic $s-$wave superconductors are not affected by the potential (non-magnetic) scattering, but are sensitive to a spin - flip scattering on magnetic impurities \cite{Anderson1959JPCS}. In high$-T_{c} $ cuprates, both magnetic and non-magnetic impurities were found to cause a rapid suppression of the superconducting transition temperature, $T_c$, strongly supporting $d-$wave pairing~\cite{Xiao1990PRB}. For an order parameter changing sign between the sheets of the Fermi surface (s$_{\pm}$ symmetry), considered the most plausible pairing state in multi-band iron-based superconductors (FeSCs)~\cite{Mazin2009PhysicaC,Hirschfeld2011ROPP,Chubukov2012ARCMP}, the response to non-magnetic scattering is expected to be significant and formally similar to the magnetic scattering in single band $s-$wave superconductors \cite{Chubukov2012ARCMP, Mishra2009PRB}. We should note that the multi-band character of superconductivity itself does not lead to the anomalous response to disorder beyond expected smearing of the gap variation on the Fermi surface, including the difference in gap magnitude between the different bands \cite{Golubov97}. For example, in a known two - gap $s_{++}$ superconductor, MgB$_2$, electron irradiation resulted only in a minor change in $T_c$ \cite{KleinPRL2010MgB2}. The sign change of the order parameter either along one sheet of the Fermi surface or between the sheets is what causes the dramatic suppression of $T_c$ due to pair-breaking nature of inter-band scattering in this case.

A relatively slow rate of $T_c$ suppression with disorder found in several experiments in iron-based superconductors was used as an argument for the $s_{{++}}$ pairing, in line with the expectations of orbital fluctuations mediated superconductivity \cite{KontaniJPSJ2008,Onari2009PRL}. In reality, the situation in sign-changing multi-band superconductors is more complicated due to the competition between intra-band and inter-band pairing and, also, band-dependent scattering times and gap anisotropies \cite{Chubukov2012ARCMP,Hirschfeld2011ROPP,WangHirschfeldMishra2013PRB}. Realistic calculations of the effect of point-like disorder on $T_c$ within $s_\pm$ scenario \cite{WangHirschfeldMishra2013PRB} agree very well with the experiment, for example in electron irradiated Ba(Fe$_{1-x}$Ru$_x$)$_2$As$_2$ \cite{Prozorov2014BaRu} and BaFe$_2$(As$_{1-x}$P$_x$)$_2$ \cite{Mizukami2014eprint}.

Experimentally, it is quite difficult to introduce controlled point - like disorder. Studies of the disorder introduced by chemical substitution~\cite{ChengWen2010PRB, Li2011PRB,Li2012sub}, heavy-ion or particle irradiation produced results that differ significantly from each other as far as the impact on the superconducting and materials properties is concerned \cite{Kim2010PRB, Tarantini2010PRL, Nakajima2010PRB, Murphy2013PRB, Salovich2013PRB,Taen2013}. Chemical substitution changes both crystalline and electronic structure and most particle irradiations introduce correlated disorder, in forms of columnar defects and/or clusters of different spatial extent \cite{Damask1963}. The effective scattering potential strength and range (size) of such defects is very different from point-like scattering. On the other hand, MeV -\ range electron irradiation is known to produce vacancy/interstitial (Frenkel) pairs, which act as efficient point - like scattering centers \cite{Damask1963}. This is consistent with the analysis of the collective pinning in  BaFe$_2$(As$_{1-x}$P$_x$)$_2$  and  Ba(Fe$_{1-x}$Co$_x$)$_2$As$_2$ \cite{VanDerBeek2013JPCS_M2S} as well as penetration depth in BaFe$_2$(As$_{1-x}$P$_x$)$_2$  \cite{Mizukami2014eprint}. In the high-\(T_{c}\) cuprates, electron irradiation defects are known to be strong unitary scatterers causing significant suppression of  $T_{c}$~\cite{Rullier-Albenque2003PRL}.

In addition to $T_c$, another parameter sensitive to disorder is quasi-particle density, which may be probed, for example, by measuring London penetration depth, $\lambda(T)$. In the case of isotropic single band $s-$wave or multiband $s_{++}-$wave superconductors, $\lambda (T)$ remains exponential at low temperatures with the addition of non-magnetic scattering \cite{TinkhamBOOK,WangHirschfeldMishra2013PRB,Mishra2009PRB}, whereas in the case of nodeless $s_{\pm}$ pairing it changes from exponential in the clean limit to $\sim T^2$ in the dirty limit~\cite{Mishra2009PRB,WangHirschfeldMishra2013PRB}. Yet, an opposite behavior is observed in superconductors with line nodes where $\lambda \sim T$ in the clean limit changing to $\sim T^2$ in the dirty limit~\cite{YipSauls1992PRL,XuYipSauls1995PRB,Hirschfeld1993PRB,KoganProzorovMishra2013PRB}. In the case of pnictide superconductors with accidental nodes, $\lambda (T)$, evolves from linear to exponential and, ultimately, to the $T^2$ behavior \cite{Mizukami2014eprint}. The details of the evolution from clean to dirty limit also depend on the (usually unknown) scattering potential amplitude and spatial extent. Weak Born scattering approximation, usually valid for normal metals, could not explain the rapid $T \rightarrow T^2$ evolution in the cuprates, so the unitary limit was used instead~\cite{KimMuzikar1994PRB, KoganProzorovMishra2013PRB,KoganProzorovMishra2013PRB}. In iron - based superconductors, the situation is unclear and it seems that intermediate scattering amplitudes (modeled within $t-$matrix approach) are needed \cite{Mishra2009PRB,WangHirschfeldMishra2013PRB,Prozorov2014BaRu}. The spatial extent of the scattering potential affects the predominant scattering $Q$-vector, and it was suggested as the cause of a notable difference in quasi-particle response and evolution of $T_c$ in SrFe$_2$(As$_{1-x}$P$_x$)$_2$ with natural and artificial disorder \cite{Strehlow2014}.

In this paper, we study the effects of electron irradiation on superconducting $T_c$ and quasi-particle excitations of hole-doped (Ba$_{1-x}$K$_x$)Fe$_2$As$_2$ single crystals by measuring in-plane resistivity, $\rho(T)$, and in-plane London penetration depth, $\Delta \lambda (T)$. (Ba$_{1-x}$K$_x$)Fe$_2$As$_2$ has the highest $T_c \approx$ 40 K among the 122 family and at the optimal doping reveals extremely robust superconductivity \cite{ReidSUST2012,KimBaK122underdoped,Salovich2013PRB}. The superconducting gap structure of (Ba$_{1-x}$K$_x$)Fe$_2$As$_2$ varies with composition from full isotropic gap at the optimal doping to the gap with line nodes at $x=$ 1 \cite{Fukazawa2009a,OkazakiOctet,KFe2As2dwave2012}. On the underdoped side of interest here, gap anisotropy increases towards the edge of the superconducting dome, especially in the range of  bulk coexistence of superconductivity and long range magnetic order \cite{KimBaK122underdoped,ZhengNMR2012}. This might imply that the ratio of the inter-band to intra-band pairing, as well as gap anisotropy, increase upon the departure from the optimal doping.

\section{Experimental}

\begin{table}
\caption{List of samples measured before and after electron irradiation. 1 C/cm$^2=$ 6.24 $\times 10^{18}$ electrons/cm$^2$.}
\label{tab.1}
\begin{center}
\begin{tabular}{|c|c|l|}
\hline
x (WDS) & sample label & measurement\\
\hline
   & 0.19-A         & $\rho$ before irradiation\\
0.19 & 0.19-A         & $\rho$ after 1.8 C/cm$^2$ irradiated\\
   & 0.19-A        & $\Delta \lambda$ after 1.8 C/cm$^2$ irradiated\\
   & 0.19-B        & $\Delta \lambda$ before irradiation\\
\hline
   & 0.26-A        & $\rho$ before irradiation\\
   & 0.26-B         & $\Delta \lambda$ before irradiation\\
0.26 & 0.26-B        & $\rho$ after 1.5 C/cm$^2$ irradiated\\
   & 0.26-B         & $\Delta \lambda$ after 1.5 C/cm$^2$ irradiated\\
   & 0.26-B         & $\Delta \lambda$ after 1.5 + 1.1 C/cm$^2$ irr.\\
\hline
   & 0.34-A        & $\rho$ before irradiation\\
   & 0.34-B         & $\Delta \lambda$ before irradiation\\
0.34 & 0.34-B        & $\rho$ after 2.0 C/cm$^2$ irradiated\\
   & 0.34-B         & $\Delta \lambda$ after 2.0 C/cm$^2$ irradiated\\
   & 0.34-C         & $\Delta \lambda$ before irradiation\\
   & 0.34-C         & $\Delta \lambda$ after 8.3 C/cm$^2$ irradiated\\
\hline
\end{tabular}
\end{center}
\end{table}

Single crystals of (Ba$_{1-x}$K$_x$)Fe$_2$As$_2$ were synthesized using high temperature FeAs flux method~\cite{Luo2008SST}. Samples used in this study were cleaved from the inner parts of single crystals and were first extensively characterized using radio-frequency magnetic susceptibility as well as magneto-optical imaging to exclude chemical and spatial inhomogeneity. The composition of each of the samples studied (see Table for the list) was measured with wavelength dispersive spectroscopy (WDS) technique, see Ref.~[\onlinecite{BaK122Tanatar2014}] for details. Four - probe measurements of in-plane resistivity were performed in {$\it Quantum$ $\it Design$} PPMS. Samples for resistivity measurements had typical dimensions of (1-2)$\times$0.5$\times$(0.02-0.1) mm$^3$. Electrical contacts to samples prior to irradiation were made by soldering 50 $\mu$m silver wires with ultra-pure tin solder, as described in Ref.~[\onlinecite{Tanatar2010SST}]. The in-plane London penetration depth, $\Delta \lambda (T)$, was measured using a self-oscillating tunnel-diode resonator technique \cite{Prozorov2006SST, Prozorov2000PRB}. The samples had typical dimensions of 0.5$\times$0.5$\times$0.03 mm$^3$. Details of the measurements and calibration can be found elsewhere \cite{Prozorov2006SST}. The 2.5 MeV electron irradiation was performed at the SIRIUS Pelletron facility of the Laboratoire des Solides Irradi\'es (LSI) at the \'Ecole Polytechnique in Palaiseau, France~\cite{VanDerBeek2013JPCS_M2S}. The dose is conveniently measured in C/cm$^2$, where 1 C/cm$^2=$ 6.24 $\times 10^{18}$ electrons/cm$^2$. The details of the measurements and doses of electron irradiation are summarized in Table~\ref{tab.1}. London penetration depth in samples 0.26-B and 0.34-B was measured in the same samples before and after the irradiation. For these samples, resistivity after electron irradiation was measured by gluing the contacts with silver paint to prevent defect annealing during soldering process. Reference samples 0.26-A and 0.34-A of the same composition for resistivity measurements were cut from the same initial larger slab. For composition $x=$ 0.19, the sample was prepared for resistivity measurements with soldered contacts. Its temperature-dependent resistivity was measured in pristine and irradiated states, see top panel of Fig.~\ref{fig.1}. After that, the contacts were removed and London penetration depth was measured.

\section{Results and Discussion}

\begin{figure}[tb]
\includegraphics[width=8.5cm]{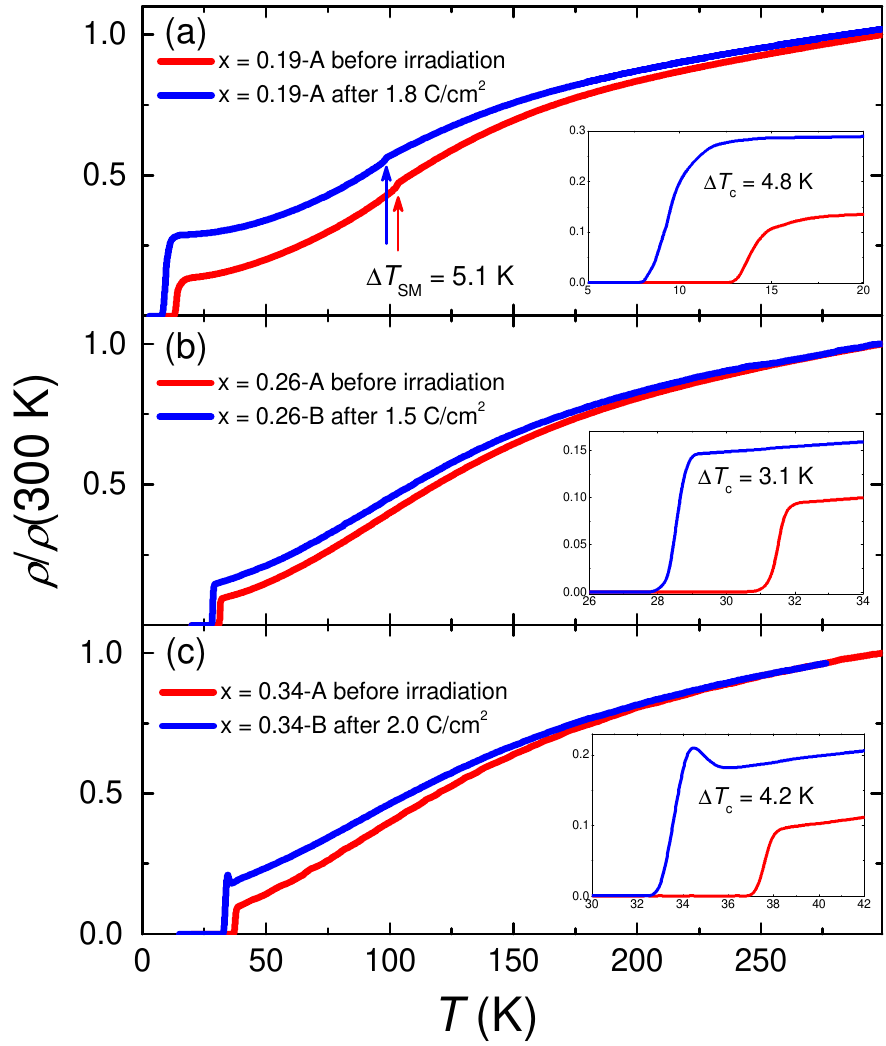}
\caption{(Color online) Temperature dependent resistivity normalized by the value at 300 K upon electron irradiation: (a) $x =$ 0.19-A, (b) $x =$ 0.26-A and B, and (c) $x=$ 0.34-A and B. Insets zoom on the superconducting transitions.}
\label{fig.1}
\end{figure}

\Fref{fig.1} shows normalized in-plane resistivity, $\rho(T)/\rho(300K)$, measured before and after electron irradiation in samples with (a) $x=$ 0.19, (b) $x=$ 0.26 and (c) $x=$ 0.34. Insets zoom on the superconducting transitions. For samples of all three doping levels, the introduction of disorder leads to a notable increase of residual  resistivity. Superconducting transition temperature, $T_c$, was determined by linear extrapolation of $\rho(T)$ at the transition to $\rho =$0.  Overall, the irradiation doses of 1.5 to 2 C/cm$^2$ lead to $T_c$ decrease by 3 to 5 K, see Fig.~\ref{fig.4}. Samples with $x =$ 0.26 and $x =$ 0.34 were outside the range of the coexisting magnetism and superconductivity.  For $x=$ 0.19, in addition to superconductivity, the long - range magnetic order develops simultaneously with orthorhombic distortion below structural/magnetic transition, $T_{sm}$. This transition is revealed as a small feature in $\rho(T)$ marked with the up-arrows in Fig.~\ref{fig.1} (a). Irradiation with 1.8 C/cm$^2$ leads to $T_{sm}$ decrease by 5.1 K, supporting the itinerant nature of antiferromagnetism \cite{TcEnhancement2012}. A ``bump" in $\rho(T)$ above $T_{c}$ developed after the irradiation in the sample with $x=$ 0.34, similar to electron irradiated YBaCuO, where it was interpreted to be due to localization effects \cite{Rullier-Albenque2003PRL}.

\begin{figure}[tb]
\includegraphics[width=8.5cm]{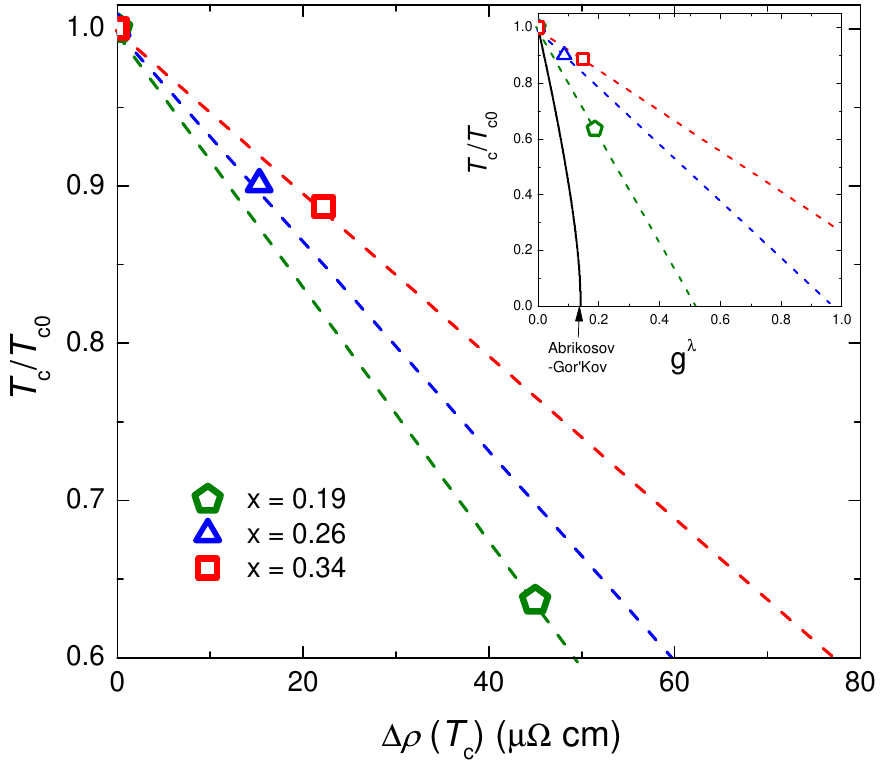}
\caption{(Color online) Experimental data of $T_c / T_{c0}$ versus $\Delta \rho (T_c)$ upon electron irradiation, where $\rho(T_c)$ is the resistivity right above $T_c$ and $\Delta \rho (T_c)$ is the change of the resistivity after irradiation. Inset shows $T_c / T_{c0}$ versus dimensionless scattering rate $g^{\lambda}$. The classical Abrikosov-Gor'kov theory for an isotropic $s-$wave superconductor with magnetic impurities is also shown by a solid line \cite{Abrikosov1960}.}
\label{fig.2}
\end{figure}

\Fref{fig.2} shows the variation of $T_c / T_{c0}$ versus $\Delta \rho_{T = T_c}$. From these values, we estimated the dimensionless scattering rate ($g^{\lambda}$) defined in a simple form as \cite{kogan2009PRB-2, Prozorov2014BaRu},

\begin{equation}
g^\lambda = \frac{\hbar}{2 \pi k_B \mu _{0}}\frac{\Delta \rho(T_c)}{T_{c0} \lambda (0) ^2}  \label{eq.1}
\end{equation}

\noindent where $\Delta \rho (T_c)$ is the change in resistivity at  $T_c$ after the irradiation. Zero - temperature London penetration depth, $\lambda (0)$, was estimated from the Homes scaling \cite{Homes2004Nature}, which takes into account both, the doping dependence and the change with pair-breaking scattering \cite{KoganProzorovMishra2013PRB}, see \fref{fig.4} and the corresponding text. The $g^\lambda$ estimated from equation \eqref{eq.1} was plotted in the inset in \fref{fig.2}. The largest variation of $d (T_c / T_{c0}) / d g^\lambda = - 1.94$ was obtained for $x =$ 0.19 while the smallest change of $d (T_c / T_{c0}) / d g^\lambda =$ - 0.76 for $x =$ 0.34. This indicates that the electron irradiation is more efficient for under-doped compounds which have a fragile superconductivity competing with long-range magnetism.

\begin{figure}[tb]
\includegraphics[width=8.5cm]{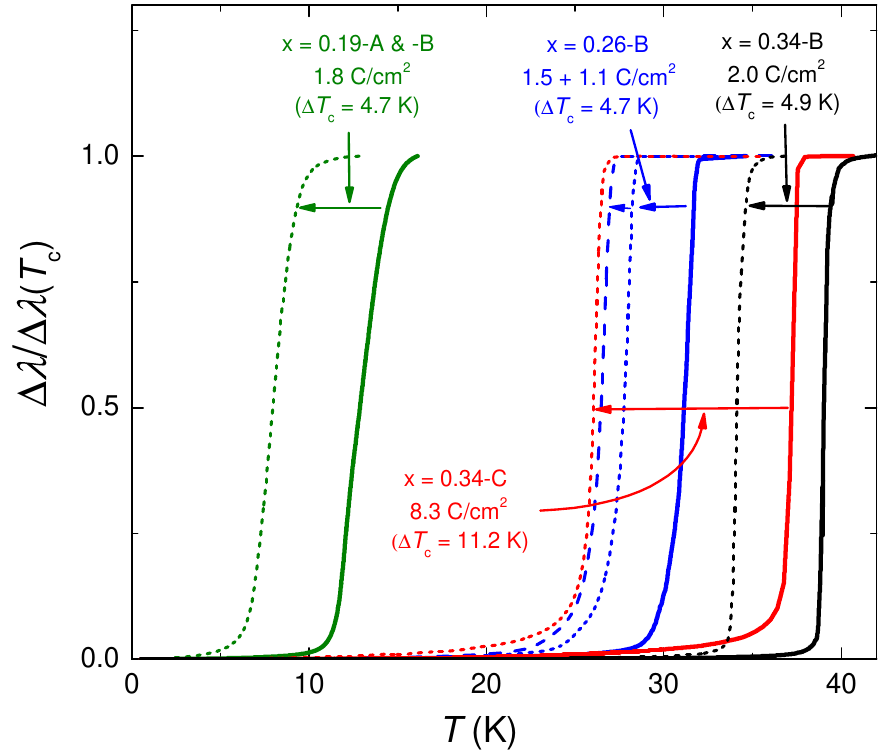}
\caption{(Color online) Variation of normalized in-plane penetration depth, $\Delta \lambda (T)/ \Delta \lambda(T_c)$, before and after electron irradiation, see Table~\ref{tab.1} for the details of the samples.}
\label{fig.3}
\end{figure}

\Fref{fig.3} shows the normalized variation of the London penetration depth, $\Delta \lambda(T)$, measured down to $\sim$ 450 mK before and after electron irradiation for all three compositions. Magnetic superconducting transition temperature, $T_c$, was defined at a point of 50$\%$ of the total change at the phase transition and was consistent with the transport measurements shown in \fref{fig.1}. The superconducting phase transition remained sharp even at the highest dose of 8.3 C/cm$^2$ that caused $T_c$ to decrease by 11.2 K for $x =$ 0.34 sample. \Fref{fig.4} summarizes the reduced $T_c$/$T_{c0}$ obtained from resistivity (open symbols) and penetration depth (full symbols) data plotted versus electron irradiation dose, where $T_{c0}$ is the $T_c$ before irradiation. The variation of $T_c$/$T_{c0}$ for $x =$ 0.34 and $x=$ 0.26 samples was about $\Delta T_c/T_{c0} \simeq$ - 0.04  per C/cm$^2$, whereas for the most underdoped sample with $x =$ 0.19 we find a five times larger value of $\Delta T_c/T_{c0} \simeq$ - 0.2 per C/cm$^2$. Quantitatively the observed doping dependence of the effect of electron irradiation on $T_c$ is similar to electron - doped Ba(Fe$_{1-x}$Co$_x$)As$_2$ ~\cite{VanDerBeek2013JPCS_M2S}.

\begin{figure}[tb]
\includegraphics[width=8.5cm]{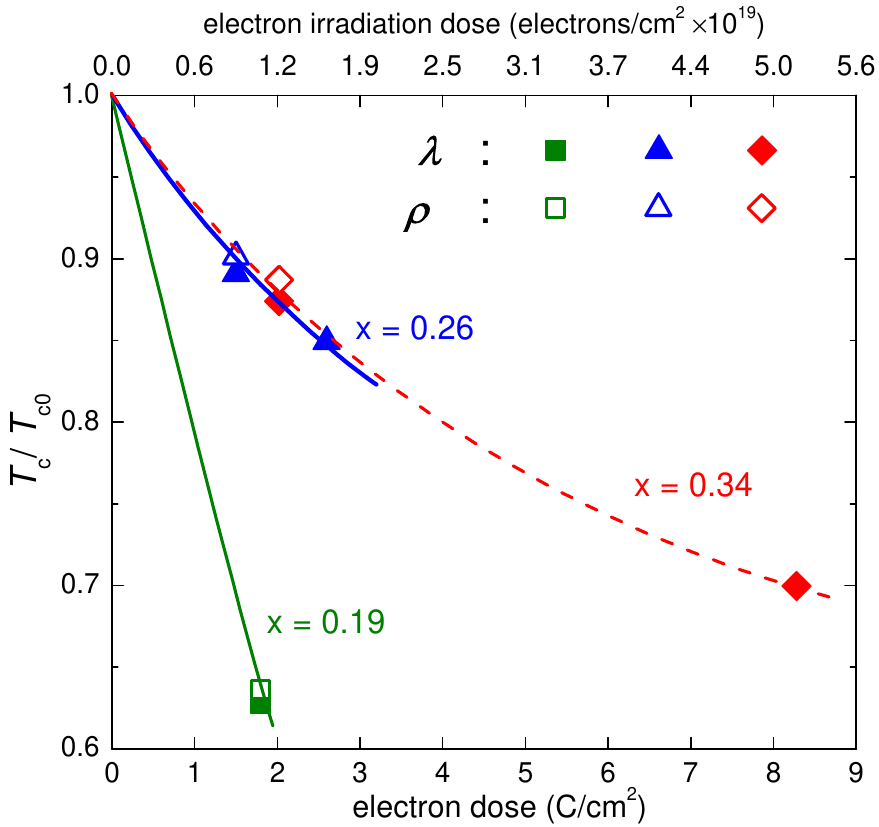}
\caption{(Color online) Variation of the reduced critical temperature, $T_c/T_{c0}$, with the dose of electron irradiation for samples $x=$ 0.19 (green squares), $x=$ 0.26 (blue triangles) and $x=$ 0.34 (red diamonds).}
\label{fig.4}
\end{figure}

\begin{figure}[tb]
\includegraphics[width=8.5cm]{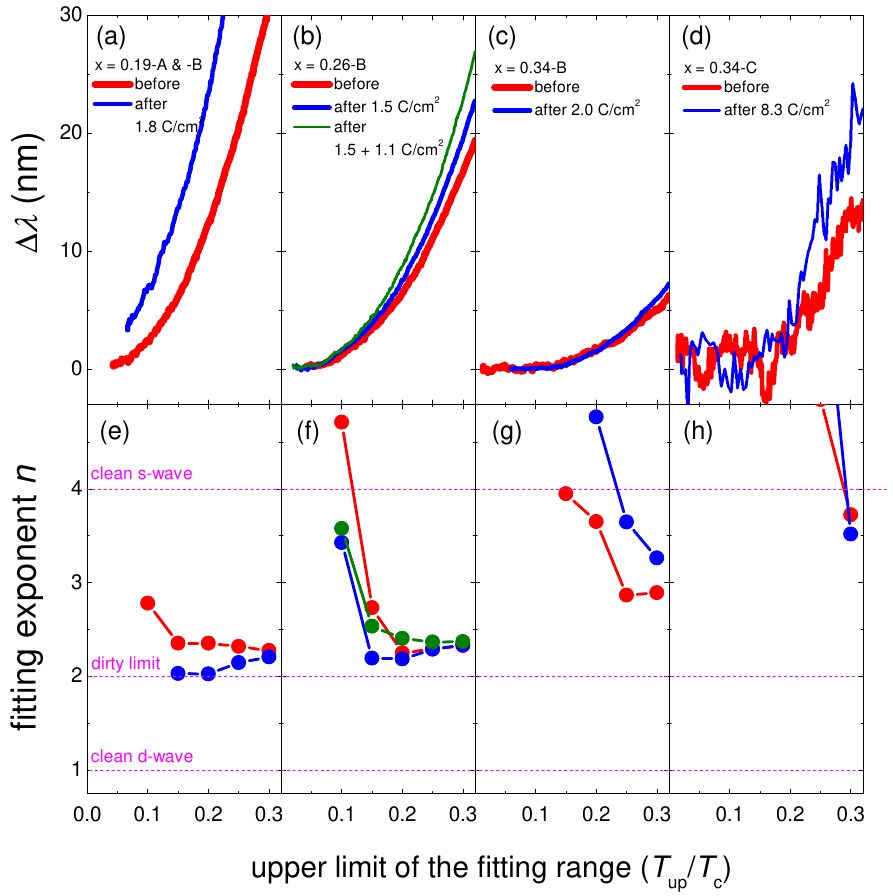}
\caption{(Color online) The low temperature part of $\Delta \lambda (T)$ before and after electron irradiation of sample $x =$ 0.19 (a), $x =$ 0.26  (b), and $x =$ 0.34  (c) and (d). The data were analyzed with a power-law function, $\Delta \lambda = A (T/T_c)^n$, over a variable temperature range from base temperature to a temperature $T_{up}$. The corresponding results of the analysis are summarized in the bottom raw of panels (e-h).}
\label{fig.5}
\end{figure}

To quantify the evolution of the superconducting gap structure, we analyzed the low temperature part of $\Delta \lambda (T)$ as shown in \fref{fig.5}. The absolute change $\Delta \lambda (T)=\lambda (0.3 T/T_c)-\lambda(T_{min}/T_c)$ clearly  increases after the irradiation indicating enhanced pair-breaking  upon introduction of additional disorder. However, the low temperature $\Delta \lambda (T)$ of the two closest to the optimal doping level samples, $x =$ 0.34-B and 0.34-C still clearly show exponential saturation below 0.2 $T/T_c$ even after the irradiation. This result suggests that the optimally - doped compositions with isotropic superconducting gaps are extremely robust against electron irradiation, even at the very large dose of 8.3 C/cm$^2$ which caused  suppression of $T_c$  by 11.2 K or $\Delta T_c \approx 0.3 T_{c0}$. The situation is similar in a slightly under-doped sample at $x =$ 0.26 in which the low temperature saturation is seen below 0.1 $T/T_c$. In a stark contrast, in the most underdoped sample with $x=$ 0.19, where superconductivity and magnetism coexist, the saturating behaviour disappears after the irradiation.

\begin{figure}[tb]
\includegraphics[width=7.0cm]{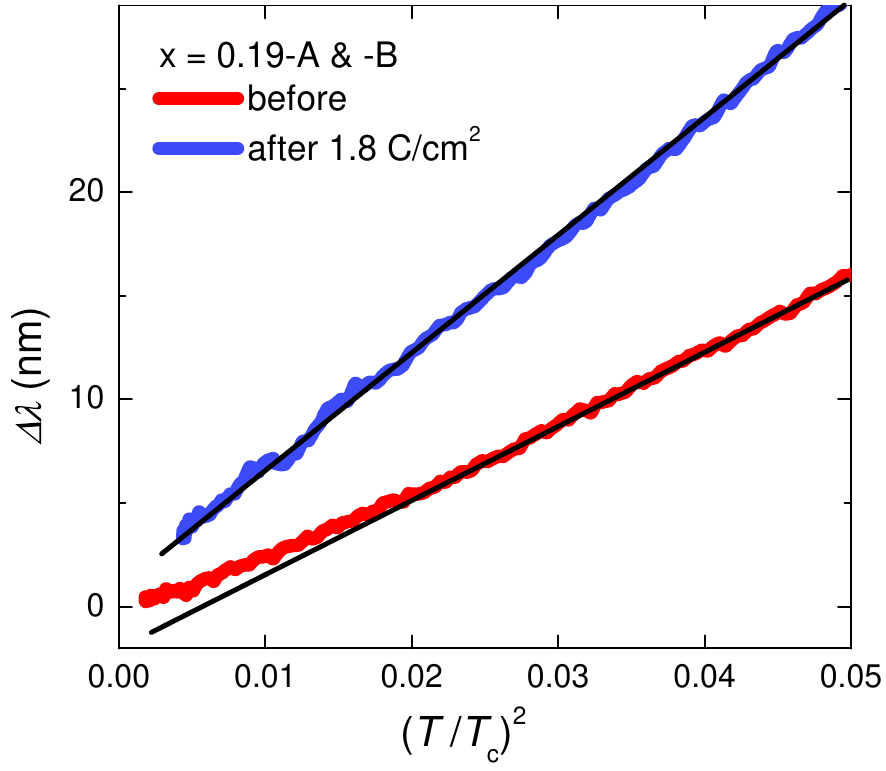}
\caption{(Color online) The low temperature part of $\Delta \lambda$ of sample $x =$ 0.19 before and after electron irradiation plotted as a function of $(T/T_c)^2$. The straight lines are the guides for the pure $T^2$ behavior.}
\label{fig.6}
\end{figure}

These observations become more apparent when the low-temperature $\Delta \lambda (T)$ is fitted using a power - law function,  $\Delta \lambda = A(T/T_c)^n$. The results are plotted in \fref{fig.5} (e-h). To eliminate the uncertainty related to the upper fitting limit, we performed several fitting runs with a variable high-temperature end of the fitting range, $T_{up}/T_c$, from 0.1 to 0.3, while keeping the lower limit at the base temperature. The strong saturation behavior of the higher - doping samples is apparent from the large exponent values, $n > 3$, even for the very large irradiation dose of 8.3 C/cm$^2$.  For $x =$ 0.26 sample, $n$ increases as the $T_{up}/T_c$ decreases. This implies that the gap remains nodeless, but develops anisotropy with a minimum value about 2 times smaller than in the case of a $x=$ 0.34 sample. For the most under-doped sample, $x =$ 0.19, there is a clear evolution toward the dirty $T^2$ limit. In the pristine state, the exponent $n$ varied from $n \approx$ 2.3 at the widest range to 2.6 - 2.8 at the narrowest $T_{up}/T_c$ range. However, after the irradiation, this tendency reverses. As the $T_{up}/T_c$ decreases, $n$ starts to decrease toward $n =$ 2. This is clearly shown in \fref{fig.6} where $\Delta \lambda$ is plotted vs. $(T/T_c)^2$. While the data before the irradiation show an upward deviation from $T^2$, the post-irradiated curve is a clean $T^2$ line.

\begin{figure}[tb]
\includegraphics[width=8.5cm]{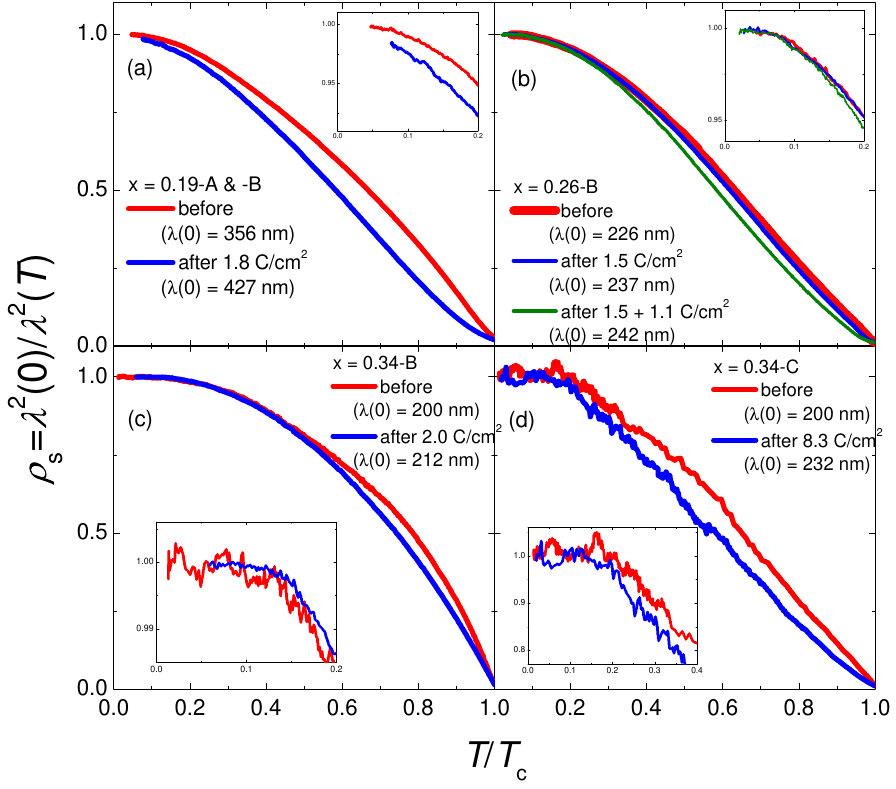}
\caption{(Color online) Normalized superfluid density, $\rho_{s} = (\lambda (0) / \lambda (T))^2$, before and after electron irradiation: for (a) $x =$ 0.19, (b) $x =$ 0.26, and (c-d) $x =$ 0.34. Doping dependent $\lambda(0)$ was estimated considering resistivity right above $T_c$ (Homes scaling) and irradiation-induced $T_c$ decrease,  see text for details.}
\label{fig.7}
\end{figure}

Another way to analyze the data is to compare the normalized superfluid density, $\rho_s(T) = \lambda^{2}(0)/\lambda^{2} (T)$. \Fref{fig.7} shows $\rho_s(T)$ before and after electron irradiation. Since the values of $\lambda (0)$ are not known, we first used the literature value of $\lambda (0)$ = 200 nm  found for unirradiated optimally doped samples, $x=$ 0.34 \cite{Martin2009PRB, RenWen2008PRL, Li2008PRL, EvtushinskyBoriesenko2009NJP}. Then, we used the Homes scaling,  $\lambda(0)\propto \rho (T_c) / T_c$ \cite{Homes2004Nature}. Here $\rho (T_c)$ is the resistivity at $T_c$. The estimated $\lambda(0)=$ 226 nm and 356 nm for $x=$ 0.26 ($T_c=$ 24.3 K) and $x=$ 0.19 ($T_c=$ 13.2 K) samples, respectively. In addition, the maximum possible increase of $\lambda(0)$ induced by the irradiation was estimated by correlating the change of $T_c$ with the pair-breaking scattering and relating it to the expected change of $ \lambda (0)$ \cite{KoganProzorovMishra2013PRB}. For example, for $x=$ 0.34-C sample, $\Delta T_c =$ - 11.2 K after 8.3 C/cm$^2$ irradiation, so the estimate of $\lambda (0)$ is about 232 nm. Following these two - step procedures, we estimated doping and irradiation dependence of $\lambda (0)$. All values are shown in the labels of \fref{fig.7}.

The superfluid density, $\rho_s(T/T_c)$, is compared in Fig.\ref{fig.7} and it is quite different for samples with different $x$. For $x =$ 0.34, panels (c-d), the overall behavior follows the expectations for an $s-$wave type pairing. Despite the change of $T_c$, the irradiation did not change the functional form of $\rho_s(T/T_c)$ much. For the more underdoped $x =$ 0.26 sample, the region of saturation shrinks, but still exists at the lowest temperatures, below 0.1 $T/T_c$. This small saturation region remains almost intact upon 2.6 C/cm$^2$ irradiation. In contrast, the superfluid density shows the largest change in the most under-doped sample, $x =$ 0.19, where even minor signs of saturation in the pre-irradiated sample disappear after the irradiation. This suggests that the superconducting gap is very anisotropic in heavily under-doped samples and, therefore, is most susceptible to the defects induced by electron irradiation. This result is also consistent with the observation that the largest $T_c$ suppression is found in the most under-doped sample, see \fref{fig.4}. Overall, the full temperature - range shape of $\rho_s(T)$ is close to a full gap $s-$wave behavior in the optimally doped sample and to a line - nodal curve for the most underdoped sample, $x =$ 0.19.

\section{Conclusions}

To summarize, the effects of electron irradiation on the in-plane resistivity and London penetration depth were studied in single crystals of hole-doped (Ba$_{1-x}$,K$_x$)Fe$_2$As$_2$ superconductor. The irradiation leads to the suppression of the superconducting $T_c$ and of the temperature of the structural/magnetic transition, $T_{sm}$. The suppression of $T_c$ is much more rapid in the underdoped sample with $x=$ 0.19, in which superconductivity coexists with the long range magnetic order. This is consistent with the development of significant gap anisotropy in the coexisting phase. In the coexisting phase, the irradiation might even induce gapless superconductivity. Considering our previous study \cite{KimBaK122underdoped} and the prediction for the rate of suppression of $T_c$ in the extended $s_\pm$ model \cite{WangHirschfeldMishra2013PRB}, we suggest that the interband to the intraband pairing ratio increases when moving away from the optimal concentration.

\section{acknowledgments}
We thank A. Chubukov, P. Hirschfeld, V. Mishra and T. Shibauchi for useful discussions. This work was supported by the U.S. Department of Energy (DOE), Office of Science, Basic Energy Sciences, Materials Science and Engineering Division. Ames Laboratory is operated for the U.S. DOE by Iowa State University under contract DE-AC02-07CH11358. We thank the SIRIUS team, B. Boizot, V. Metayer, and J. Losco, for running electron irradiation at \textit{Ecole Polytechnique} (supported by EMIR network, proposal 11-11-0121.) Work in China was supported by the Ministry of Science and Technology of China, project 2011CBA00102.


\begin{thebibliography}{99}%
\makeatletter
\providecommand \@ifxundefined [1]{%
 \@ifx{#1\undefined}
}%
\providecommand \@ifnum [1]{%
 \ifnum #1\expandafter \@firstoftwo
 \else \expandafter \@secondoftwo
 \fi
}%
\providecommand \@ifx [1]{%
 \ifx #1\expandafter \@firstoftwo
 \else \expandafter \@secondoftwo
 \fi
}%
\providecommand \natexlab [1]{#1}%
\providecommand \enquote  [1]{``#1''}%
\providecommand \bibnamefont  [1]{#1}%
\providecommand \bibfnamefont [1]{#1}%
\providecommand \citenamefont [1]{#1}%
\providecommand \href@noop [0]{\@secondoftwo}%
\providecommand \href [0]{\begingroup \@sanitize@url \@href}%
\providecommand \@href[1]{\@@startlink{#1}\@@href}%
\providecommand \@@href[1]{\endgroup#1\@@endlink}%
\providecommand \@sanitize@url [0]{\catcode `\\12\catcode `\$12\catcode
  `\&12\catcode `\#12\catcode `\^12\catcode `\_12\catcode `\%12\relax}%
\providecommand \@@startlink[1]{}%
\providecommand \@@endlink[0]{}%
\providecommand \url  [0]{\begingroup\@sanitize@url \@url }%
\providecommand \@url [1]{\endgroup\@href {#1}{\urlprefix }}%
\providecommand \urlprefix  [0]{URL }%
\providecommand \Eprint [0]{\href }%
\providecommand \doibase [0]{http://dx.doi.org/}%
\providecommand \selectlanguage [0]{\@gobble}%
\providecommand \bibinfo  [0]{\@secondoftwo}%
\providecommand \bibfield  [0]{\@secondoftwo}%
\providecommand \translation [1]{[#1]}%
\providecommand \BibitemOpen [0]{}%
\providecommand \bibitemStop [0]{}%
\providecommand \bibitemNoStop [0]{.\EOS\space}%
\providecommand \EOS [0]{\spacefactor3000\relax}%
\providecommand \BibitemShut  [1]{\csname bibitem#1\endcsname}%
\let\auto@bib@innerbib\@empty
\bibitem [{\citenamefont {Anderson}(1959)}]{Anderson1959JPCS}%
  \BibitemOpen
  \bibfield  {author} {\bibinfo {author} {\bibfnamefont {P.~W.}\ \bibnamefont
  {Anderson}},\ }\href@noop {} {\bibfield  {journal} {\bibinfo  {journal} {J
  Phys Chem Solids}\ }\textbf {\bibinfo {volume} {11}},\ \bibinfo {pages} {26}
  (\bibinfo {year} {1959})}\BibitemShut {NoStop}%
\bibitem [{\citenamefont {Abrikosov}\ and\ \citenamefont
  {Gor'kov}(1960)}]{Abrikosov1960}%
  \BibitemOpen
  \bibfield  {author} {\bibinfo {author} {\bibfnamefont {A.~A.}\ \bibnamefont
  {Abrikosov}}\ and\ \bibinfo {author} {\bibfnamefont {L.~P.}\ \bibnamefont
  {Gor'kov}},\ }\href@noop {} {\bibfield  {journal} {\bibinfo  {journal} {Zh.
  Eksp. Teor. Fiz. [Sov. Phys. JETP 12, 1243 (1961)]}\ }\textbf {\bibinfo
  {volume} {39}},\ \bibinfo {pages} {1781} (\bibinfo {year}
  {1960})}\BibitemShut {NoStop}%
\bibitem [{\citenamefont {Hirschfeld}\ and\ \citenamefont
  {Goldenfeld}(1993)}]{Hirschfeld1993PRB}%
  \BibitemOpen
  \bibfield  {author} {\bibinfo {author} {\bibfnamefont {P.~J.}\ \bibnamefont
  {Hirschfeld}}\ and\ \bibinfo {author} {\bibfnamefont {N.}~\bibnamefont
  {Goldenfeld}},\ }\href@noop {} {\bibfield  {journal} {\bibinfo  {journal}
  {Phys. Rev. B}\ }\textbf {\bibinfo {volume} {48}},\ \bibinfo {pages} {4219}
  (\bibinfo {year} {1993})}\BibitemShut {NoStop}%
\bibitem [{\citenamefont {Kim}\ \emph {et~al.}(1994)\citenamefont {Kim},
  \citenamefont {Preosti},\ and\ \citenamefont {Muzikar}}]{KimMuzikar1994PRB}%
  \BibitemOpen
  \bibfield  {author} {\bibinfo {author} {\bibfnamefont {H.}~\bibnamefont
  {Kim}}, \bibinfo {author} {\bibfnamefont {G.}~\bibnamefont {Preosti}}, \ and\
  \bibinfo {author} {\bibfnamefont {P.}~\bibnamefont {Muzikar}},\ }\href@noop
  {} {\bibfield  {journal} {\bibinfo  {journal} {Phys. Rev. B}\ }\textbf
  {\bibinfo {volume} {49}},\ \bibinfo {pages} {3544} (\bibinfo {year}
  {1994})}\BibitemShut {NoStop}%
\bibitem [{\citenamefont {Balatsky}\ \emph {et~al.}(2006)\citenamefont
  {Balatsky}, \citenamefont {Vekhter},\ and\ \citenamefont
  {Zhu}}]{Balatsky2006RMP}%
  \BibitemOpen
  \bibfield  {author} {\bibinfo {author} {\bibfnamefont {A.~V.}\ \bibnamefont
  {Balatsky}}, \bibinfo {author} {\bibfnamefont {I.}~\bibnamefont {Vekhter}}, \
  and\ \bibinfo {author} {\bibfnamefont {J.-X.}\ \bibnamefont {Zhu}},\
  }\href@noop {} {\bibfield  {journal} {\bibinfo  {journal} {Rev. Mod. Phys.}\
  }\textbf {\bibinfo {volume} {78}},\ \bibinfo {pages} {373} (\bibinfo {year}
  {2006})}\BibitemShut {NoStop}%
\bibitem [{\citenamefont {Wang}\ \emph {et~al.}(2013)\citenamefont {Wang},
  \citenamefont {Kreisel}, \citenamefont {Hirschfeld},\ and\ \citenamefont
  {Mishra}}]{WangHirschfeldMishra2013PRB}%
  \BibitemOpen
  \bibfield  {author} {\bibinfo {author} {\bibfnamefont {Y.}~\bibnamefont
  {Wang}}, \bibinfo {author} {\bibfnamefont {A.}~\bibnamefont {Kreisel}},
  \bibinfo {author} {\bibfnamefont {P.~J.}\ \bibnamefont {Hirschfeld}}, \ and\
  \bibinfo {author} {\bibfnamefont {V.}~\bibnamefont {Mishra}},\ }\href@noop {}
  {\bibfield  {journal} {\bibinfo  {journal} {Phys. Rev. B}\ }\textbf {\bibinfo
  {volume} {87}},\ \bibinfo {pages} {094504} (\bibinfo {year}
  {2013})}\BibitemShut {NoStop}%
\bibitem [{\citenamefont {Prozorov}\ \emph {et~al.}(2014)\citenamefont
  {Prozorov}, \citenamefont {Kończykowski}, \citenamefont {Tanatar},
  \citenamefont {Thaler}, \citenamefont {Bud'ko}, \citenamefont {Canfield},
  \citenamefont {Mishra},\ and\ \citenamefont {Hirschfeld}}]{Prozorov2014BaRu}%
  \BibitemOpen
  \bibfield  {author} {\bibinfo {author} {\bibfnamefont {R.}~\bibnamefont
  {Prozorov}}, \bibinfo {author} {\bibfnamefont {M.}~\bibnamefont
  {Kończykowski}}, \bibinfo {author} {\bibfnamefont {M.~A.}\ \bibnamefont
  {Tanatar}}, \bibinfo {author} {\bibfnamefont {A.}~\bibnamefont {Thaler}},
  \bibinfo {author} {\bibfnamefont {S.~L.}\ \bibnamefont {Bud'ko}}, \bibinfo
  {author} {\bibfnamefont {P.~C.}\ \bibnamefont {Canfield}}, \bibinfo {author}
  {\bibfnamefont {V.}~\bibnamefont {Mishra}}, \ and\ \bibinfo {author}
  {\bibfnamefont {P.~J.}\ \bibnamefont {Hirschfeld}},\ }\href@noop {}
  {\bibfield  {journal} {\bibinfo  {journal} {arXiv:1405.3255}\ } (\bibinfo
  {year} {2014})},\ \Eprint {http://arxiv.org/abs/1405.3255} {1405.3255}
  \BibitemShut {NoStop}%
\bibitem [{\citenamefont {Mizukami}\ \emph {et~al.}(2014)\citenamefont
  {Mizukami}, \citenamefont {Konczykowski}, \citenamefont {Kawamoto},
  \citenamefont {Kurata}, \citenamefont {Kasahara}, \citenamefont {Hashimoto},
  \citenamefont {Mishra}, \citenamefont {Kreisel}, \citenamefont {Wang},
  \citenamefont {Hirschfeld}, \citenamefont {Matsuda},\ and\ \citenamefont
  {Shibauchi}}]{Mizukami2014eprint}%
  \BibitemOpen
  \bibfield  {author} {\bibinfo {author} {\bibfnamefont {Y.}~\bibnamefont
  {Mizukami}}, \bibinfo {author} {\bibfnamefont {M.}~\bibnamefont
  {Konczykowski}}, \bibinfo {author} {\bibfnamefont {Y.}~\bibnamefont
  {Kawamoto}}, \bibinfo {author} {\bibfnamefont {S.}~\bibnamefont {Kurata}},
  \bibinfo {author} {\bibfnamefont {S.}~\bibnamefont {Kasahara}}, \bibinfo
  {author} {\bibfnamefont {K.}~\bibnamefont {Hashimoto}}, \bibinfo {author}
  {\bibfnamefont {V.}~\bibnamefont {Mishra}}, \bibinfo {author} {\bibfnamefont
  {A.}~\bibnamefont {Kreisel}}, \bibinfo {author} {\bibfnamefont
  {Y.}~\bibnamefont {Wang}}, \bibinfo {author} {\bibfnamefont {P.~J.}\
  \bibnamefont {Hirschfeld}}, \bibinfo {author} {\bibfnamefont
  {Y.}~\bibnamefont {Matsuda}}, \ and\ \bibinfo {author} {\bibfnamefont
  {T.}~\bibnamefont {Shibauchi}},\ }\href@noop {} {\bibfield  {journal}
  {\bibinfo  {journal} {arXiv:1405.6951}\ } (\bibinfo {year}
  {2014})}\BibitemShut {NoStop}%
\bibitem [{\citenamefont {Xiao}\ \emph {et~al.}(1990)\citenamefont {Xiao},
  \citenamefont {Cieplak}, \citenamefont {Xiao},\ and\ \citenamefont
  {Chien}}]{Xiao1990PRB}%
  \BibitemOpen
  \bibfield  {author} {\bibinfo {author} {\bibfnamefont {G.}~\bibnamefont
  {Xiao}}, \bibinfo {author} {\bibfnamefont {M.~Z.}\ \bibnamefont {Cieplak}},
  \bibinfo {author} {\bibfnamefont {J.~Q.}\ \bibnamefont {Xiao}}, \ and\
  \bibinfo {author} {\bibfnamefont {C.~L.}\ \bibnamefont {Chien}},\ }\href@noop
  {} {\bibfield  {journal} {\bibinfo  {journal} {Phys. Rev. B}\ }\textbf
  {\bibinfo {volume} {42}},\ \bibinfo {pages} {8752} (\bibinfo {year}
  {1990})}\BibitemShut {NoStop}%
\bibitem [{\citenamefont {Mazin}\ and\ \citenamefont
  {Schmalian}(2009)}]{Mazin2009PhysicaC}%
  \BibitemOpen
  \bibfield  {author} {\bibinfo {author} {\bibfnamefont {I.}~\bibnamefont
  {Mazin}}\ and\ \bibinfo {author} {\bibfnamefont {J.}~\bibnamefont
  {Schmalian}},\ }\href@noop {} {\bibfield  {journal} {\bibinfo  {journal}
  {Physica C}\ }\textbf {\bibinfo {volume} {469}},\ \bibinfo {pages} {614}
  (\bibinfo {year} {2009})}\BibitemShut {NoStop}%
\bibitem [{\citenamefont {Hirschfeld}\ \emph {et~al.}(2011)\citenamefont
  {Hirschfeld}, \citenamefont {Korshunov},\ and\ \citenamefont
  {Mazin}}]{Hirschfeld2011ROPP}%
  \BibitemOpen
  \bibfield  {author} {\bibinfo {author} {\bibfnamefont {P.~J.}\ \bibnamefont
  {Hirschfeld}}, \bibinfo {author} {\bibfnamefont {M.~M.}\ \bibnamefont
  {Korshunov}}, \ and\ \bibinfo {author} {\bibfnamefont {I.~I.}\ \bibnamefont
  {Mazin}},\ }\href {http://stacks.iop.org/0034-4885/74/i=12/a=124508}
  {\bibfield  {journal} {\bibinfo  {journal} {Rep Prog Phys}\ }\textbf
  {\bibinfo {volume} {74}},\ \bibinfo {pages} {124508} (\bibinfo {year}
  {2011})}\BibitemShut {NoStop}%
\bibitem [{\citenamefont {Chubukov}(2012)}]{Chubukov2012ARCMP}%
  \BibitemOpen
  \bibfield  {author} {\bibinfo {author} {\bibfnamefont {A.}~\bibnamefont
  {Chubukov}},\ }\href@noop {} {\bibfield  {journal} {\bibinfo  {journal} {Ann.
  Rev. Cond. Matt. Phys.}\ }\textbf {\bibinfo {volume} {3}},\ \bibinfo {pages}
  {13} (\bibinfo {year} {2012})}\BibitemShut {NoStop}%
\bibitem [{\citenamefont {Mishra}\ \emph {et~al.}(2009)\citenamefont {Mishra},
  \citenamefont {Boyd}, \citenamefont {Graser}, \citenamefont {Maier},
  \citenamefont {Hirschfeld},\ and\ \citenamefont {Scalapino}}]{Mishra2009PRB}%
  \BibitemOpen
  \bibfield  {author} {\bibinfo {author} {\bibfnamefont {V.}~\bibnamefont
  {Mishra}}, \bibinfo {author} {\bibfnamefont {G.}~\bibnamefont {Boyd}},
  \bibinfo {author} {\bibfnamefont {S.}~\bibnamefont {Graser}}, \bibinfo
  {author} {\bibfnamefont {T.}~\bibnamefont {Maier}}, \bibinfo {author}
  {\bibfnamefont {P.~J.}\ \bibnamefont {Hirschfeld}}, \ and\ \bibinfo {author}
  {\bibfnamefont {D.~J.}\ \bibnamefont {Scalapino}},\ }\href@noop {} {\bibfield
   {journal} {\bibinfo  {journal} {Phys. Rev. B}\ }\textbf {\bibinfo {volume}
  {79}},\ \bibinfo {pages} {094512} (\bibinfo {year} {2009})}\BibitemShut
  {NoStop}%
\bibitem [{\citenamefont {Golubov}\ and\ \citenamefont
  {Mazin}(1997)}]{Golubov97}%
  \BibitemOpen
  \bibfield  {author} {\bibinfo {author} {\bibfnamefont {A.~A.}\ \bibnamefont
  {Golubov}}\ and\ \bibinfo {author} {\bibfnamefont {I.~I.}\ \bibnamefont
  {Mazin}},\ }\href {\doibase 10.1103/PhysRevB.55.15146} {\bibfield  {journal}
  {\bibinfo  {journal} {Phys. Rev. B}\ }\textbf {\bibinfo {volume} {55}},\
  \bibinfo {pages} {15146} (\bibinfo {year} {1997})}\BibitemShut {NoStop}%
\bibitem [{\citenamefont {Klein}\ \emph {et~al.}(2010)\citenamefont {Klein},
  \citenamefont {Marlaud}, \citenamefont {Marcenat}, \citenamefont
  {Cercellier}, \citenamefont {Konczykowski}, \citenamefont {van~der Beek},
  \citenamefont {Mosser}, \citenamefont {Lee},\ and\ \citenamefont
  {Lee}}]{KleinPRL2010MgB2}%
  \BibitemOpen
  \bibfield  {author} {\bibinfo {author} {\bibfnamefont {T.}~\bibnamefont
  {Klein}}, \bibinfo {author} {\bibfnamefont {R.}~\bibnamefont {Marlaud}},
  \bibinfo {author} {\bibfnamefont {C.}~\bibnamefont {Marcenat}}, \bibinfo
  {author} {\bibfnamefont {H.}~\bibnamefont {Cercellier}}, \bibinfo {author}
  {\bibfnamefont {M.}~\bibnamefont {Konczykowski}}, \bibinfo {author}
  {\bibfnamefont {C.~J.}\ \bibnamefont {van~der Beek}}, \bibinfo {author}
  {\bibfnamefont {V.}~\bibnamefont {Mosser}}, \bibinfo {author} {\bibfnamefont
  {H.~S.}\ \bibnamefont {Lee}}, \ and\ \bibinfo {author} {\bibfnamefont
  {S.~I.}\ \bibnamefont {Lee}},\ }\href {\doibase
  10.1103/PhysRevLett.105.047001} {\bibfield  {journal} {\bibinfo  {journal}
  {Phys Rev Lett}\ }\textbf {\bibinfo {volume} {105}},\ \bibinfo {pages}
  {047001} (\bibinfo {year} {2010})}\BibitemShut {NoStop}%
\bibitem [{\citenamefont {Senga}\ and\ \citenamefont
  {Kontani}(2008)}]{KontaniJPSJ2008}%
  \BibitemOpen
  \bibfield  {author} {\bibinfo {author} {\bibfnamefont {Y.}~\bibnamefont
  {Senga}}\ and\ \bibinfo {author} {\bibfnamefont {H.}~\bibnamefont
  {Kontani}},\ }\href {\doibase 10.1143/JPSJ.77.113710} {\bibfield  {journal}
  {\bibinfo  {journal} {J. Phys. Soc. Jpn.}\ }\textbf {\bibinfo {volume}
  {77}},\ \bibinfo {pages} {113710} (\bibinfo {year} {2008})}\BibitemShut
  {NoStop}%
\bibitem [{\citenamefont {Onari}\ and\ \citenamefont
  {Kontani}(2009)}]{Onari2009PRL}%
  \BibitemOpen
  \bibfield  {author} {\bibinfo {author} {\bibfnamefont {S.}~\bibnamefont
  {Onari}}\ and\ \bibinfo {author} {\bibfnamefont {H.}~\bibnamefont
  {Kontani}},\ }\href@noop {} {\bibfield  {journal} {\bibinfo  {journal} {Phys
  Rev Lett}\ }\textbf {\bibinfo {volume} {103}},\ \bibinfo {pages} {177001}
  (\bibinfo {year} {2009})}\BibitemShut {NoStop}%
\bibitem [{\citenamefont {Cheng}\ \emph {et~al.}(2010)\citenamefont {Cheng},
  \citenamefont {Shen}, \citenamefont {Hu},\ and\ \citenamefont
  {Wen}}]{ChengWen2010PRB}%
  \BibitemOpen
  \bibfield  {author} {\bibinfo {author} {\bibfnamefont {P.}~\bibnamefont
  {Cheng}}, \bibinfo {author} {\bibfnamefont {B.}~\bibnamefont {Shen}},
  \bibinfo {author} {\bibfnamefont {J.}~\bibnamefont {Hu}}, \ and\ \bibinfo
  {author} {\bibfnamefont {H.-H.}\ \bibnamefont {Wen}},\ }\href@noop {}
  {\bibfield  {journal} {\bibinfo  {journal} {Phys. Rev. B}\ }\textbf {\bibinfo
  {volume} {81}},\ \bibinfo {pages} {174529} (\bibinfo {year}
  {2010})}\BibitemShut {NoStop}%
\bibitem [{\citenamefont {Li}\ \emph {et~al.}(2011)\citenamefont {Li},
  \citenamefont {Guo}, \citenamefont {Zhang}, \citenamefont {Yu}, \citenamefont
  {Tsujimoto}, \citenamefont {Kontani}, \citenamefont {Yamaura},\ and\
  \citenamefont {Takayama-Muromachi}}]{Li2011PRB}%
  \BibitemOpen
  \bibfield  {author} {\bibinfo {author} {\bibfnamefont {J.}~\bibnamefont
  {Li}}, \bibinfo {author} {\bibfnamefont {Y.}~\bibnamefont {Guo}}, \bibinfo
  {author} {\bibfnamefont {S.}~\bibnamefont {Zhang}}, \bibinfo {author}
  {\bibfnamefont {S.}~\bibnamefont {Yu}}, \bibinfo {author} {\bibfnamefont
  {Y.}~\bibnamefont {Tsujimoto}}, \bibinfo {author} {\bibfnamefont
  {H.}~\bibnamefont {Kontani}}, \bibinfo {author} {\bibfnamefont
  {K.}~\bibnamefont {Yamaura}}, \ and\ \bibinfo {author} {\bibfnamefont
  {E.}~\bibnamefont {Takayama-Muromachi}},\ }\href@noop {} {\bibfield
  {journal} {\bibinfo  {journal} {Phys. Rev. B}\ }\textbf {\bibinfo {volume}
  {84}},\ \bibinfo {pages} {020513} (\bibinfo {year} {2011})}\BibitemShut
  {NoStop}%
\bibitem [{\citenamefont {Li}\ \emph {et~al.}(2012{\natexlab{a}})\citenamefont
  {Li}, \citenamefont {Guo}, \citenamefont {Zhang}, \citenamefont {Yuan},
  \citenamefont {Tsujimoto}, \citenamefont {Wang}, \citenamefont {Sathish},
  \citenamefont {Sun}, \citenamefont {Yu}, \citenamefont {Yi}, \citenamefont
  {Yamaura}, \citenamefont {Takayama-Muromachiu}, \citenamefont {Shirako},
  \citenamefont {Akaogi},\ and\ \citenamefont {Kontani}}]{Li2012sub}%
  \BibitemOpen
  \bibfield  {author} {\bibinfo {author} {\bibfnamefont {J.}~\bibnamefont
  {Li}}, \bibinfo {author} {\bibfnamefont {Y.~F.}\ \bibnamefont {Guo}},
  \bibinfo {author} {\bibfnamefont {S.~B.}\ \bibnamefont {Zhang}}, \bibinfo
  {author} {\bibfnamefont {J.}~\bibnamefont {Yuan}}, \bibinfo {author}
  {\bibfnamefont {Y.}~\bibnamefont {Tsujimoto}}, \bibinfo {author}
  {\bibfnamefont {X.}~\bibnamefont {Wang}}, \bibinfo {author} {\bibfnamefont
  {C.~I.}\ \bibnamefont {Sathish}}, \bibinfo {author} {\bibfnamefont
  {Y.}~\bibnamefont {Sun}}, \bibinfo {author} {\bibfnamefont {S.}~\bibnamefont
  {Yu}}, \bibinfo {author} {\bibfnamefont {W.}~\bibnamefont {Yi}}, \bibinfo
  {author} {\bibfnamefont {K.}~\bibnamefont {Yamaura}}, \bibinfo {author}
  {\bibfnamefont {E.}~\bibnamefont {Takayama-Muromachiu}}, \bibinfo {author}
  {\bibfnamefont {Y.}~\bibnamefont {Shirako}}, \bibinfo {author} {\bibfnamefont
  {M.}~\bibnamefont {Akaogi}}, \ and\ \bibinfo {author} {\bibfnamefont
  {H.}~\bibnamefont {Kontani}},\ }\href@noop {} {\bibfield  {journal} {\bibinfo
   {journal} {Phys. Rev. B}\ }\textbf {\bibinfo {volume} {85}},\ \bibinfo
  {pages} {214509} (\bibinfo {year} {2012}{\natexlab{a}})}\BibitemShut
  {NoStop}%
\bibitem [{\citenamefont {Kim}\ \emph {et~al.}(2010)\citenamefont {Kim},
  \citenamefont {Gordon}, \citenamefont {Tanatar}, \citenamefont {Hua},
  \citenamefont {Welp}, \citenamefont {Kwok}, \citenamefont {Ni}, \citenamefont
  {Bud'ko}, \citenamefont {Canfield}, \citenamefont {Vorontsov},\ and\
  \citenamefont {Prozorov}}]{Kim2010PRB}%
  \BibitemOpen
  \bibfield  {author} {\bibinfo {author} {\bibfnamefont {H.}~\bibnamefont
  {Kim}}, \bibinfo {author} {\bibfnamefont {R.~T.}\ \bibnamefont {Gordon}},
  \bibinfo {author} {\bibfnamefont {M.~A.}\ \bibnamefont {Tanatar}}, \bibinfo
  {author} {\bibfnamefont {J.}~\bibnamefont {Hua}}, \bibinfo {author}
  {\bibfnamefont {U.}~\bibnamefont {Welp}}, \bibinfo {author} {\bibfnamefont
  {W.~K.}\ \bibnamefont {Kwok}}, \bibinfo {author} {\bibfnamefont
  {N.}~\bibnamefont {Ni}}, \bibinfo {author} {\bibfnamefont {S.~L.}\
  \bibnamefont {Bud'ko}}, \bibinfo {author} {\bibfnamefont {P.~C.}\
  \bibnamefont {Canfield}}, \bibinfo {author} {\bibfnamefont {A.~B.}\
  \bibnamefont {Vorontsov}}, \ and\ \bibinfo {author} {\bibfnamefont
  {R.}~\bibnamefont {Prozorov}},\ }\href@noop {} {\bibfield  {journal}
  {\bibinfo  {journal} {Phys. Rev. B}\ }\textbf {\bibinfo {volume} {82}},\
  \bibinfo {pages} {060518} (\bibinfo {year} {2010})}\BibitemShut {NoStop}%
\bibitem [{\citenamefont {Tarantini}\ \emph {et~al.}(2010)\citenamefont
  {Tarantini}, \citenamefont {Putti}, \citenamefont {Gurevich}, \citenamefont
  {Shen}, \citenamefont {Singh}, \citenamefont {Rowell}, \citenamefont
  {Newman}, \citenamefont {Larbalestier}, \citenamefont {Cheng}, \citenamefont
  {Jia},\ and\ \citenamefont {Wen}}]{Tarantini2010PRL}%
  \BibitemOpen
  \bibfield  {author} {\bibinfo {author} {\bibfnamefont {C.}~\bibnamefont
  {Tarantini}}, \bibinfo {author} {\bibfnamefont {M.}~\bibnamefont {Putti}},
  \bibinfo {author} {\bibfnamefont {A.}~\bibnamefont {Gurevich}}, \bibinfo
  {author} {\bibfnamefont {Y.}~\bibnamefont {Shen}}, \bibinfo {author}
  {\bibfnamefont {R.~K.}\ \bibnamefont {Singh}}, \bibinfo {author}
  {\bibfnamefont {J.~M.}\ \bibnamefont {Rowell}}, \bibinfo {author}
  {\bibfnamefont {N.}~\bibnamefont {Newman}}, \bibinfo {author} {\bibfnamefont
  {D.~C.}\ \bibnamefont {Larbalestier}}, \bibinfo {author} {\bibfnamefont
  {P.}~\bibnamefont {Cheng}}, \bibinfo {author} {\bibfnamefont
  {Y.}~\bibnamefont {Jia}}, \ and\ \bibinfo {author} {\bibfnamefont {H.-H.}\
  \bibnamefont {Wen}},\ }\href@noop {} {\bibfield  {journal} {\bibinfo
  {journal} {Phys Rev Lett}\ }\textbf {\bibinfo {volume} {104}},\ \bibinfo
  {pages} {087002} (\bibinfo {year} {2010})}\BibitemShut {NoStop}%
\bibitem [{\citenamefont {Nakajima}\ \emph {et~al.}(2010)\citenamefont
  {Nakajima}, \citenamefont {Taen}, \citenamefont {Tsuchiya}, \citenamefont
  {Tamegai}, \citenamefont {Kitamura},\ and\ \citenamefont
  {Murakami}}]{Nakajima2010PRB}%
  \BibitemOpen
  \bibfield  {author} {\bibinfo {author} {\bibfnamefont {Y.}~\bibnamefont
  {Nakajima}}, \bibinfo {author} {\bibfnamefont {T.}~\bibnamefont {Taen}},
  \bibinfo {author} {\bibfnamefont {Y.}~\bibnamefont {Tsuchiya}}, \bibinfo
  {author} {\bibfnamefont {T.}~\bibnamefont {Tamegai}}, \bibinfo {author}
  {\bibfnamefont {H.}~\bibnamefont {Kitamura}}, \ and\ \bibinfo {author}
  {\bibfnamefont {T.}~\bibnamefont {Murakami}},\ }\href@noop {} {\bibfield
  {journal} {\bibinfo  {journal} {Phys. Rev. B}\ }\textbf {\bibinfo {volume}
  {82}},\ \bibinfo {pages} {220504} (\bibinfo {year} {2010})}\BibitemShut
  {NoStop}%
\bibitem [{\citenamefont {Murphy}\ \emph {et~al.}(2013)\citenamefont {Murphy},
  \citenamefont {Tanatar}, \citenamefont {Kim}, \citenamefont {Kwok},
  \citenamefont {Welp}, \citenamefont {Graf}, \citenamefont {Brooks},
  \citenamefont {Bud'ko}, \citenamefont {Canfield},\ and\ \citenamefont
  {Prozorov}}]{Murphy2013PRB}%
  \BibitemOpen
  \bibfield  {author} {\bibinfo {author} {\bibfnamefont {J.}~\bibnamefont
  {Murphy}}, \bibinfo {author} {\bibfnamefont {M.~A.}\ \bibnamefont {Tanatar}},
  \bibinfo {author} {\bibfnamefont {H.}~\bibnamefont {Kim}}, \bibinfo {author}
  {\bibfnamefont {W.}~\bibnamefont {Kwok}}, \bibinfo {author} {\bibfnamefont
  {U.}~\bibnamefont {Welp}}, \bibinfo {author} {\bibfnamefont {D.}~\bibnamefont
  {Graf}}, \bibinfo {author} {\bibfnamefont {J.~S.}\ \bibnamefont {Brooks}},
  \bibinfo {author} {\bibfnamefont {S.~L.}\ \bibnamefont {Bud'ko}}, \bibinfo
  {author} {\bibfnamefont {P.~C.}\ \bibnamefont {Canfield}}, \ and\ \bibinfo
  {author} {\bibfnamefont {R.}~\bibnamefont {Prozorov}},\ }\href@noop {}
  {\bibfield  {journal} {\bibinfo  {journal} {Phys. Rev. B}\ }\textbf {\bibinfo
  {volume} {88}},\ \bibinfo {pages} {054514} (\bibinfo {year}
  {2013})}\BibitemShut {NoStop}%
\bibitem [{\citenamefont {Salovich}\ \emph {et~al.}(2013)\citenamefont
  {Salovich}, \citenamefont {Kim}, \citenamefont {Ghosh}, \citenamefont
  {Giannetta}, \citenamefont {Kwok}, \citenamefont {Welp}, \citenamefont
  {Shen}, \citenamefont {Zhu}, \citenamefont {Wen}, \citenamefont {Tanatar},\
  and\ \citenamefont {Prozorov}}]{Salovich2013PRB}%
  \BibitemOpen
  \bibfield  {author} {\bibinfo {author} {\bibfnamefont {N.~W.}\ \bibnamefont
  {Salovich}}, \bibinfo {author} {\bibfnamefont {H.}~\bibnamefont {Kim}},
  \bibinfo {author} {\bibfnamefont {A.~K.}\ \bibnamefont {Ghosh}}, \bibinfo
  {author} {\bibfnamefont {R.~W.}\ \bibnamefont {Giannetta}}, \bibinfo {author}
  {\bibfnamefont {W.}~\bibnamefont {Kwok}}, \bibinfo {author} {\bibfnamefont
  {U.}~\bibnamefont {Welp}}, \bibinfo {author} {\bibfnamefont {B.}~\bibnamefont
  {Shen}}, \bibinfo {author} {\bibfnamefont {S.}~\bibnamefont {Zhu}}, \bibinfo
  {author} {\bibfnamefont {H.-H.}\ \bibnamefont {Wen}}, \bibinfo {author}
  {\bibfnamefont {M.~A.}\ \bibnamefont {Tanatar}}, \ and\ \bibinfo {author}
  {\bibfnamefont {R.}~\bibnamefont {Prozorov}},\ }\href@noop {} {\bibfield
  {journal} {\bibinfo  {journal} {Phys. Rev. B}\ }\textbf {\bibinfo {volume}
  {87}},\ \bibinfo {pages} {180502} (\bibinfo {year} {2013})}\BibitemShut
  {NoStop}%
\bibitem [{\citenamefont {Taen}\ \emph {et~al.}(2013)\citenamefont {Taen},
  \citenamefont {Ohtake}, \citenamefont {Akiyama}, \citenamefont {Inoue},
  \citenamefont {Sun}, \citenamefont {Pyon}, \citenamefont {Tamegai},\ and\
  \citenamefont {Kitamura}}]{Taen2013}%
  \BibitemOpen
  \bibfield  {author} {\bibinfo {author} {\bibfnamefont {T.}~\bibnamefont
  {Taen}}, \bibinfo {author} {\bibfnamefont {F.}~\bibnamefont {Ohtake}},
  \bibinfo {author} {\bibfnamefont {H.}~\bibnamefont {Akiyama}}, \bibinfo
  {author} {\bibfnamefont {H.}~\bibnamefont {Inoue}}, \bibinfo {author}
  {\bibfnamefont {Y.}~\bibnamefont {Sun}}, \bibinfo {author} {\bibfnamefont
  {S.}~\bibnamefont {Pyon}}, \bibinfo {author} {\bibfnamefont {T.}~\bibnamefont
  {Tamegai}}, \ and\ \bibinfo {author} {\bibfnamefont {H.}~\bibnamefont
  {Kitamura}},\ }\href@noop {} {\bibfield  {journal} {\bibinfo  {journal}
  {Phys. Rev. B}\ }\textbf {\bibinfo {volume} {88}},\ \bibinfo {pages} {224514}
  (\bibinfo {year} {2013})}\BibitemShut {NoStop}%
\bibitem [{\citenamefont {Damask}\ and\ \citenamefont
  {Dienes}(1963)}]{Damask1963}%
  \BibitemOpen
  \bibfield  {author} {\bibinfo {author} {\bibfnamefont {A.~C.}\ \bibnamefont
  {Damask}}\ and\ \bibinfo {author} {\bibfnamefont {G.~J.}\ \bibnamefont
  {Dienes}},\ }\href@noop {} {\emph {\bibinfo {title} {Point Defects in
  Metals}}}\ (\bibinfo  {publisher} {Gordon and Breach Science Publishers
  Ltd},\ \bibinfo {year} {1963})\BibitemShut {NoStop}%
\bibitem [{\citenamefont {van~der Beek}\ \emph {et~al.}(2013)\citenamefont
  {van~der Beek}, \citenamefont {Demirdis}, \citenamefont {Colson},
  \citenamefont {Rullier-Albenque}, \citenamefont {Fasano}, \citenamefont
  {Shibauchi}, \citenamefont {Matsuda}, \citenamefont {Kasahara}, \citenamefont
  {Gierlowski},\ and\ \citenamefont {Konczykowski}}]{VanDerBeek2013JPCS_M2S}%
  \BibitemOpen
  \bibfield  {author} {\bibinfo {author} {\bibfnamefont {C.~J.}\ \bibnamefont
  {van~der Beek}}, \bibinfo {author} {\bibfnamefont {S.}~\bibnamefont
  {Demirdis}}, \bibinfo {author} {\bibfnamefont {D.}~\bibnamefont {Colson}},
  \bibinfo {author} {\bibfnamefont {F.}~\bibnamefont {Rullier-Albenque}},
  \bibinfo {author} {\bibfnamefont {Y.}~\bibnamefont {Fasano}}, \bibinfo
  {author} {\bibfnamefont {T.}~\bibnamefont {Shibauchi}}, \bibinfo {author}
  {\bibfnamefont {Y.}~\bibnamefont {Matsuda}}, \bibinfo {author} {\bibfnamefont
  {S.}~\bibnamefont {Kasahara}}, \bibinfo {author} {\bibfnamefont
  {P.}~\bibnamefont {Gierlowski}}, \ and\ \bibinfo {author} {\bibfnamefont
  {M.}~\bibnamefont {Konczykowski}},\ }\href@noop {} {\bibfield  {journal}
  {\bibinfo  {journal} {Journal of Physics: Conference Series}\ }\textbf
  {\bibinfo {volume} {449}},\ \bibinfo {pages} {012023} (\bibinfo {year}
  {2013})}\BibitemShut {NoStop}%
\bibitem [{\citenamefont {Rullier-Albenque}\ \emph {et~al.}(2003)\citenamefont
  {Rullier-Albenque}, \citenamefont {Alloul},\ and\ \citenamefont
  {Tourbot}}]{Rullier-Albenque2003PRL}%
  \BibitemOpen
  \bibfield  {author} {\bibinfo {author} {\bibfnamefont {F.}~\bibnamefont
  {Rullier-Albenque}}, \bibinfo {author} {\bibfnamefont {H.}~\bibnamefont
  {Alloul}}, \ and\ \bibinfo {author} {\bibfnamefont {R.}~\bibnamefont
  {Tourbot}},\ }\href@noop {} {\bibfield  {journal} {\bibinfo  {journal} {Phys
  Rev Lett}\ }\textbf {\bibinfo {volume} {91}},\ \bibinfo {pages} {047001}
  (\bibinfo {year} {2003})}\BibitemShut {NoStop}%
\bibitem [{\citenamefont {Tinkham}(2004)}]{TinkhamBOOK}%
  \BibitemOpen
  \bibfield  {author} {\bibinfo {author} {\bibfnamefont {M.}~\bibnamefont
  {Tinkham}},\ }\href@noop {} {\emph {\bibinfo {title} {Introduction to
  Superconductivity}}},\ \bibinfo {edition} {2nd}\ ed.,\ Dover Books on
  Physics\ (\bibinfo  {publisher} {Dover Publications},\ \bibinfo {year}
  {2004})\BibitemShut {NoStop}%
\bibitem [{\citenamefont {Yip}\ and\ \citenamefont
  {Sauls}(1992)}]{YipSauls1992PRL}%
  \BibitemOpen
  \bibfield  {author} {\bibinfo {author} {\bibfnamefont {S.~K.}\ \bibnamefont
  {Yip}}\ and\ \bibinfo {author} {\bibfnamefont {J.~A.}\ \bibnamefont
  {Sauls}},\ }\href@noop {} {\bibfield  {journal} {\bibinfo  {journal} {Phys
  Rev Lett}\ }\textbf {\bibinfo {volume} {69}},\ \bibinfo {pages} {2264}
  (\bibinfo {year} {1992})}\BibitemShut {NoStop}%
\bibitem [{\citenamefont {Xu}\ \emph {et~al.}(1995)\citenamefont {Xu},
  \citenamefont {Yip},\ and\ \citenamefont {Sauls}}]{XuYipSauls1995PRB}%
  \BibitemOpen
  \bibfield  {author} {\bibinfo {author} {\bibfnamefont {D.}~\bibnamefont
  {Xu}}, \bibinfo {author} {\bibfnamefont {S.~K.}\ \bibnamefont {Yip}}, \ and\
  \bibinfo {author} {\bibfnamefont {J.~A.}\ \bibnamefont {Sauls}},\ }\href@noop
  {} {\bibfield  {journal} {\bibinfo  {journal} {Phys. Rev. B}\ }\textbf
  {\bibinfo {volume} {51}},\ \bibinfo {pages} {16233} (\bibinfo {year}
  {1995})}\BibitemShut {NoStop}%
\bibitem [{\citenamefont {Kogan}\ \emph {et~al.}(2013)\citenamefont {Kogan},
  \citenamefont {Prozorov},\ and\ \citenamefont
  {Mishra}}]{KoganProzorovMishra2013PRB}%
  \BibitemOpen
  \bibfield  {author} {\bibinfo {author} {\bibfnamefont {V.~G.}\ \bibnamefont
  {Kogan}}, \bibinfo {author} {\bibfnamefont {R.}~\bibnamefont {Prozorov}}, \
  and\ \bibinfo {author} {\bibfnamefont {V.}~\bibnamefont {Mishra}},\
  }\href@noop {} {\bibfield  {journal} {\bibinfo  {journal} {Phys. Rev. B}\
  }\textbf {\bibinfo {volume} {88}},\ \bibinfo {pages} {224508} (\bibinfo
  {year} {2013})}\BibitemShut {NoStop}%
\bibitem [{\citenamefont {Strehlow}\ \emph {et~al.}(2014)\citenamefont
  {Strehlow}, \citenamefont {Ko\ifmmode~\acute{n}\else \'{n}\fi{}czykowski},
  \citenamefont {Murphy}, \citenamefont {Teknowijoyo}, \citenamefont {Cho},
  \citenamefont {Tanatar}, \citenamefont {Kobayashi}, \citenamefont {Miyasaka},
  \citenamefont {Tajima},\ and\ \citenamefont {Prozorov}}]{Strehlow2014}%
  \BibitemOpen
  \bibfield  {author} {\bibinfo {author} {\bibfnamefont {C.~P.}\ \bibnamefont
  {Strehlow}}, \bibinfo {author} {\bibfnamefont {M.}~\bibnamefont
  {Ko\ifmmode~\acute{n}\else \'{n}\fi{}czykowski}}, \bibinfo {author}
  {\bibfnamefont {J.~A.}\ \bibnamefont {Murphy}}, \bibinfo {author}
  {\bibfnamefont {S.}~\bibnamefont {Teknowijoyo}}, \bibinfo {author}
  {\bibfnamefont {K.}~\bibnamefont {Cho}}, \bibinfo {author} {\bibfnamefont
  {M.~A.}\ \bibnamefont {Tanatar}}, \bibinfo {author} {\bibfnamefont
  {T.}~\bibnamefont {Kobayashi}}, \bibinfo {author} {\bibfnamefont
  {S.}~\bibnamefont {Miyasaka}}, \bibinfo {author} {\bibfnamefont
  {S.}~\bibnamefont {Tajima}}, \ and\ \bibinfo {author} {\bibfnamefont
  {R.}~\bibnamefont {Prozorov}},\ }\href {\doibase 10.1103/PhysRevB.90.020508}
  {\bibfield  {journal} {\bibinfo  {journal} {Phys. Rev. B}\ }\textbf {\bibinfo
  {volume} {90}},\ \bibinfo {pages} {020508} (\bibinfo {year}
  {2014})}\BibitemShut {NoStop}%
\bibitem [{\citenamefont {Reid}\ \emph
  {et~al.}(2012{\natexlab{a}})\citenamefont {Reid}, \citenamefont
  {Juneau-Fecteau}, \citenamefont {Gordon}, \citenamefont {de~Cotret},
  \citenamefont {Doiron-Leyraud}, \citenamefont {Luo}, \citenamefont
  {Shakeripour}, \citenamefont {Chang}, \citenamefont {Tanatar}, \citenamefont
  {Kim}, \citenamefont {Prozorov}, \citenamefont {Saito}, \citenamefont
  {Fukazawa}, \citenamefont {Kohori}, \citenamefont {Kihou}, \citenamefont
  {Lee}, \citenamefont {Iyo}, \citenamefont {Eisaki}, \citenamefont {Shen},
  \citenamefont {Wen},\ and\ \citenamefont {Taillefer}}]{ReidSUST2012}%
  \BibitemOpen
  \bibfield  {author} {\bibinfo {author} {\bibfnamefont {J.-P.}\ \bibnamefont
  {Reid}}, \bibinfo {author} {\bibfnamefont {A.}~\bibnamefont
  {Juneau-Fecteau}}, \bibinfo {author} {\bibfnamefont {R.~T.}\ \bibnamefont
  {Gordon}}, \bibinfo {author} {\bibfnamefont {S.~R.}\ \bibnamefont
  {de~Cotret}}, \bibinfo {author} {\bibfnamefont {N.}~\bibnamefont
  {Doiron-Leyraud}}, \bibinfo {author} {\bibfnamefont {X.~G.}\ \bibnamefont
  {Luo}}, \bibinfo {author} {\bibfnamefont {H.}~\bibnamefont {Shakeripour}},
  \bibinfo {author} {\bibfnamefont {J.}~\bibnamefont {Chang}}, \bibinfo
  {author} {\bibfnamefont {M.~A.}\ \bibnamefont {Tanatar}}, \bibinfo {author}
  {\bibfnamefont {H.}~\bibnamefont {Kim}}, \bibinfo {author} {\bibfnamefont
  {R.}~\bibnamefont {Prozorov}}, \bibinfo {author} {\bibfnamefont
  {T.}~\bibnamefont {Saito}}, \bibinfo {author} {\bibfnamefont
  {H.}~\bibnamefont {Fukazawa}}, \bibinfo {author} {\bibfnamefont
  {Y.}~\bibnamefont {Kohori}}, \bibinfo {author} {\bibfnamefont
  {K.}~\bibnamefont {Kihou}}, \bibinfo {author} {\bibfnamefont {C.~H.}\
  \bibnamefont {Lee}}, \bibinfo {author} {\bibfnamefont {A.}~\bibnamefont
  {Iyo}}, \bibinfo {author} {\bibfnamefont {H.}~\bibnamefont {Eisaki}},
  \bibinfo {author} {\bibfnamefont {B.}~\bibnamefont {Shen}}, \bibinfo {author}
  {\bibfnamefont {H.-H.}\ \bibnamefont {Wen}}, \ and\ \bibinfo {author}
  {\bibfnamefont {L.}~\bibnamefont {Taillefer}},\ }\href
  {http://stacks.iop.org/0953-2048/25/i=8/a=084013} {\bibfield  {journal}
  {\bibinfo  {journal} {Superconductor Science and Technology}\ }\textbf
  {\bibinfo {volume} {25}},\ \bibinfo {pages} {084013} (\bibinfo {year}
  {2012}{\natexlab{a}})}\BibitemShut {NoStop}%
\bibitem [{\citenamefont {Kim}\ \emph {et~al.}(2014)\citenamefont {Kim},
  \citenamefont {Tanatar}, \citenamefont {Straszheim}, \citenamefont {Cho},
  \citenamefont {Murphy}, \citenamefont {Spyrison}, \citenamefont {Reid},
  \citenamefont {Shen}, \citenamefont {Wen}, \citenamefont {Fernandes},\ and\
  \citenamefont {Prozorov}}]{KimBaK122underdoped}%
  \BibitemOpen
  \bibfield  {author} {\bibinfo {author} {\bibfnamefont {H.}~\bibnamefont
  {Kim}}, \bibinfo {author} {\bibfnamefont {M.~A.}\ \bibnamefont {Tanatar}},
  \bibinfo {author} {\bibfnamefont {W.~E.}\ \bibnamefont {Straszheim}},
  \bibinfo {author} {\bibfnamefont {K.}~\bibnamefont {Cho}}, \bibinfo {author}
  {\bibfnamefont {J.}~\bibnamefont {Murphy}}, \bibinfo {author} {\bibfnamefont
  {N.}~\bibnamefont {Spyrison}}, \bibinfo {author} {\bibfnamefont {J.-P.}\
  \bibnamefont {Reid}}, \bibinfo {author} {\bibfnamefont {B.}~\bibnamefont
  {Shen}}, \bibinfo {author} {\bibfnamefont {H.-H.}\ \bibnamefont {Wen}},
  \bibinfo {author} {\bibfnamefont {R.~M.}\ \bibnamefont {Fernandes}}, \ and\
  \bibinfo {author} {\bibfnamefont {R.}~\bibnamefont {Prozorov}},\ }\href
  {\doibase 10.1103/PhysRevB.90.014517} {\bibfield  {journal} {\bibinfo
  {journal} {Phys. Rev. B}\ }\textbf {\bibinfo {volume} {90}},\ \bibinfo
  {pages} {014517} (\bibinfo {year} {2014})}\BibitemShut {NoStop}%
\bibitem [{\citenamefont {Fukazawa}\ \emph {et~al.}(2009)\citenamefont
  {Fukazawa}, \citenamefont {Yamada}, \citenamefont {Kondo}, \citenamefont
  {Saito}, \citenamefont {Kohori}, \citenamefont {Kuga}, \citenamefont
  {Matsumoto}, \citenamefont {Nakatsuji}, \citenamefont {Kito}, \citenamefont
  {Shirage}, \citenamefont {Kihou}, \citenamefont {Takeshita}, \citenamefont
  {Lee}, \citenamefont {Iyo},\ and\ \citenamefont {Eisaki}}]{Fukazawa2009a}%
  \BibitemOpen
  \bibfield  {author} {\bibinfo {author} {\bibfnamefont {H.}~\bibnamefont
  {Fukazawa}}, \bibinfo {author} {\bibfnamefont {Y.}~\bibnamefont {Yamada}},
  \bibinfo {author} {\bibfnamefont {K.}~\bibnamefont {Kondo}}, \bibinfo
  {author} {\bibfnamefont {T.}~\bibnamefont {Saito}}, \bibinfo {author}
  {\bibfnamefont {Y.}~\bibnamefont {Kohori}}, \bibinfo {author} {\bibfnamefont
  {K.}~\bibnamefont {Kuga}}, \bibinfo {author} {\bibfnamefont {Y.}~\bibnamefont
  {Matsumoto}}, \bibinfo {author} {\bibfnamefont {S.}~\bibnamefont
  {Nakatsuji}}, \bibinfo {author} {\bibfnamefont {H.}~\bibnamefont {Kito}},
  \bibinfo {author} {\bibfnamefont {P.~M.}\ \bibnamefont {Shirage}}, \bibinfo
  {author} {\bibfnamefont {K.}~\bibnamefont {Kihou}}, \bibinfo {author}
  {\bibfnamefont {N.}~\bibnamefont {Takeshita}}, \bibinfo {author}
  {\bibfnamefont {C.-H.}\ \bibnamefont {Lee}}, \bibinfo {author} {\bibfnamefont
  {A.}~\bibnamefont {Iyo}}, \ and\ \bibinfo {author} {\bibfnamefont
  {H.}~\bibnamefont {Eisaki}},\ }\href {\doibase 10.1143/JPSJ.78.083712}
  {\bibfield  {journal} {\bibinfo  {journal} {Journal of the Physical Society
  of Japan}\ }\textbf {\bibinfo {volume} {78}},\ \bibinfo {pages} {083712}
  (\bibinfo {year} {2009})}\BibitemShut {NoStop}%
\bibitem [{\citenamefont {Okazaki}\ \emph {et~al.}(2012)\citenamefont
  {Okazaki}, \citenamefont {Ota}, \citenamefont {Kotani}, \citenamefont
  {Malaeb}, \citenamefont {Ishida}, \citenamefont {Shimojima}, \citenamefont
  {Kiss}, \citenamefont {Watanabe}, \citenamefont {Chen}, \citenamefont
  {Kihou}, \citenamefont {Lee}, \citenamefont {Iyo}, \citenamefont {Eisaki},
  \citenamefont {Saito}, \citenamefont {Fukazawa}, \citenamefont {Kohori},
  \citenamefont {Hashimoto}, \citenamefont {Shibauchi}, \citenamefont
  {Matsuda}, \citenamefont {Ikeda}, \citenamefont {Miyahara}, \citenamefont
  {Arita}, \citenamefont {Chainani},\ and\ \citenamefont
  {Shin}}]{OkazakiOctet}%
  \BibitemOpen
  \bibfield  {author} {\bibinfo {author} {\bibfnamefont {K.}~\bibnamefont
  {Okazaki}}, \bibinfo {author} {\bibfnamefont {Y.}~\bibnamefont {Ota}},
  \bibinfo {author} {\bibfnamefont {Y.}~\bibnamefont {Kotani}}, \bibinfo
  {author} {\bibfnamefont {W.}~\bibnamefont {Malaeb}}, \bibinfo {author}
  {\bibfnamefont {Y.}~\bibnamefont {Ishida}}, \bibinfo {author} {\bibfnamefont
  {T.}~\bibnamefont {Shimojima}}, \bibinfo {author} {\bibfnamefont
  {T.}~\bibnamefont {Kiss}}, \bibinfo {author} {\bibfnamefont {S.}~\bibnamefont
  {Watanabe}}, \bibinfo {author} {\bibfnamefont {C.-T.}\ \bibnamefont {Chen}},
  \bibinfo {author} {\bibfnamefont {K.}~\bibnamefont {Kihou}}, \bibinfo
  {author} {\bibfnamefont {C.~H.}\ \bibnamefont {Lee}}, \bibinfo {author}
  {\bibfnamefont {A.}~\bibnamefont {Iyo}}, \bibinfo {author} {\bibfnamefont
  {H.}~\bibnamefont {Eisaki}}, \bibinfo {author} {\bibfnamefont
  {T.}~\bibnamefont {Saito}}, \bibinfo {author} {\bibfnamefont
  {H.}~\bibnamefont {Fukazawa}}, \bibinfo {author} {\bibfnamefont
  {Y.}~\bibnamefont {Kohori}}, \bibinfo {author} {\bibfnamefont
  {K.}~\bibnamefont {Hashimoto}}, \bibinfo {author} {\bibfnamefont
  {T.}~\bibnamefont {Shibauchi}}, \bibinfo {author} {\bibfnamefont
  {Y.}~\bibnamefont {Matsuda}}, \bibinfo {author} {\bibfnamefont
  {H.}~\bibnamefont {Ikeda}}, \bibinfo {author} {\bibfnamefont
  {H.}~\bibnamefont {Miyahara}}, \bibinfo {author} {\bibfnamefont
  {R.}~\bibnamefont {Arita}}, \bibinfo {author} {\bibfnamefont
  {A.}~\bibnamefont {Chainani}}, \ and\ \bibinfo {author} {\bibfnamefont
  {S.}~\bibnamefont {Shin}},\ }\href@noop {} {\bibfield  {journal} {\bibinfo
  {journal} {Science}\ }\textbf {\bibinfo {volume} {337}},\ \bibinfo {pages}
  {1314} (\bibinfo {year} {2012})}\BibitemShut {NoStop}%
\bibitem [{\citenamefont {Reid}\ \emph
  {et~al.}(2012{\natexlab{b}})\citenamefont {Reid}, \citenamefont {Tanatar},
  \citenamefont {Juneau-Fecteau}, \citenamefont {Gordon}, \citenamefont
  {de~Cotret}, \citenamefont {Doiron-Leyraud}, \citenamefont {Saito},
  \citenamefont {Fukazawa}, \citenamefont {Kohori}, \citenamefont {Kihou},
  \citenamefont {Lee}, \citenamefont {Iyo}, \citenamefont {Eisaki},
  \citenamefont {Prozorov},\ and\ \citenamefont
  {Taillefer}}]{KFe2As2dwave2012}%
  \BibitemOpen
  \bibfield  {author} {\bibinfo {author} {\bibfnamefont {J.-P.}\ \bibnamefont
  {Reid}}, \bibinfo {author} {\bibfnamefont {M.~A.}\ \bibnamefont {Tanatar}},
  \bibinfo {author} {\bibfnamefont {A.}~\bibnamefont {Juneau-Fecteau}},
  \bibinfo {author} {\bibfnamefont {R.~T.}\ \bibnamefont {Gordon}}, \bibinfo
  {author} {\bibfnamefont {S.~R.}\ \bibnamefont {de~Cotret}}, \bibinfo {author}
  {\bibfnamefont {N.}~\bibnamefont {Doiron-Leyraud}}, \bibinfo {author}
  {\bibfnamefont {T.}~\bibnamefont {Saito}}, \bibinfo {author} {\bibfnamefont
  {H.}~\bibnamefont {Fukazawa}}, \bibinfo {author} {\bibfnamefont
  {Y.}~\bibnamefont {Kohori}}, \bibinfo {author} {\bibfnamefont
  {K.}~\bibnamefont {Kihou}}, \bibinfo {author} {\bibfnamefont {C.~H.}\
  \bibnamefont {Lee}}, \bibinfo {author} {\bibfnamefont {A.}~\bibnamefont
  {Iyo}}, \bibinfo {author} {\bibfnamefont {H.}~\bibnamefont {Eisaki}},
  \bibinfo {author} {\bibfnamefont {R.}~\bibnamefont {Prozorov}}, \ and\
  \bibinfo {author} {\bibfnamefont {L.}~\bibnamefont {Taillefer}},\ }\href
  {\doibase 10.1103/PhysRevLett.109.087001} {\bibfield  {journal} {\bibinfo
  {journal} {Phys. Rev. Lett.}\ }\textbf {\bibinfo {volume} {109}},\ \bibinfo
  {pages} {087001} (\bibinfo {year} {2012}{\natexlab{b}})}\BibitemShut
  {NoStop}%
\bibitem [{\citenamefont {Li}\ \emph {et~al.}(2012{\natexlab{b}})\citenamefont
  {Li}, \citenamefont {Zhou}, \citenamefont {Liu}, \citenamefont {Sun},
  \citenamefont {Yang}, \citenamefont {Lin},\ and\ \citenamefont
  {Zheng}}]{ZhengNMR2012}%
  \BibitemOpen
  \bibfield  {author} {\bibinfo {author} {\bibfnamefont {Z.}~\bibnamefont
  {Li}}, \bibinfo {author} {\bibfnamefont {R.}~\bibnamefont {Zhou}}, \bibinfo
  {author} {\bibfnamefont {Y.}~\bibnamefont {Liu}}, \bibinfo {author}
  {\bibfnamefont {D.~L.}\ \bibnamefont {Sun}}, \bibinfo {author} {\bibfnamefont
  {J.}~\bibnamefont {Yang}}, \bibinfo {author} {\bibfnamefont {C.~T.}\
  \bibnamefont {Lin}}, \ and\ \bibinfo {author} {\bibfnamefont {G.-q.}\
  \bibnamefont {Zheng}},\ }\href {\doibase 10.1103/PhysRevB.86.180501}
  {\bibfield  {journal} {\bibinfo  {journal} {Phys. Rev. B}\ }\textbf {\bibinfo
  {volume} {86}},\ \bibinfo {pages} {180501} (\bibinfo {year}
  {2012}{\natexlab{b}})}\BibitemShut {NoStop}%
\bibitem [{\citenamefont {Luo}\ \emph {et~al.}(2008)\citenamefont {Luo},
  \citenamefont {Wang}, \citenamefont {Yang}, \citenamefont {Cheng},
  \citenamefont {Zhu},\ and\ \citenamefont {Wen}}]{Luo2008SST}%
  \BibitemOpen
  \bibfield  {author} {\bibinfo {author} {\bibfnamefont {H.}~\bibnamefont
  {Luo}}, \bibinfo {author} {\bibfnamefont {Z.}~\bibnamefont {Wang}}, \bibinfo
  {author} {\bibfnamefont {H.}~\bibnamefont {Yang}}, \bibinfo {author}
  {\bibfnamefont {P.}~\bibnamefont {Cheng}}, \bibinfo {author} {\bibfnamefont
  {X.}~\bibnamefont {Zhu}}, \ and\ \bibinfo {author} {\bibfnamefont {H.-H.}\
  \bibnamefont {Wen}},\ }\href@noop {} {\bibfield  {journal} {\bibinfo
  {journal} {Supercond. Sci. Technol.}\ }\textbf {\bibinfo {volume} {21}},\
  \bibinfo {pages} {125014} (\bibinfo {year} {2008})}\BibitemShut {NoStop}%
\bibitem [{\citenamefont {Tanatar}\ \emph {et~al.}(2014)\citenamefont
  {Tanatar}, \citenamefont {Straszheim}, \citenamefont {Kim}, \citenamefont
  {Murphy}, \citenamefont {Spyrison}, \citenamefont {Blomberg}, \citenamefont
  {Cho}, \citenamefont {Reid}, \citenamefont {Shen}, \citenamefont {Taillefer},
  \citenamefont {Wen},\ and\ \citenamefont {Prozorov}}]{BaK122Tanatar2014}%
  \BibitemOpen
  \bibfield  {author} {\bibinfo {author} {\bibfnamefont {M.~A.}\ \bibnamefont
  {Tanatar}}, \bibinfo {author} {\bibfnamefont {W.~E.}\ \bibnamefont
  {Straszheim}}, \bibinfo {author} {\bibfnamefont {H.}~\bibnamefont {Kim}},
  \bibinfo {author} {\bibfnamefont {J.}~\bibnamefont {Murphy}}, \bibinfo
  {author} {\bibfnamefont {N.}~\bibnamefont {Spyrison}}, \bibinfo {author}
  {\bibfnamefont {E.~C.}\ \bibnamefont {Blomberg}}, \bibinfo {author}
  {\bibfnamefont {K.}~\bibnamefont {Cho}}, \bibinfo {author} {\bibfnamefont
  {J.-P.}\ \bibnamefont {Reid}}, \bibinfo {author} {\bibfnamefont
  {B.}~\bibnamefont {Shen}}, \bibinfo {author} {\bibfnamefont {L.}~\bibnamefont
  {Taillefer}}, \bibinfo {author} {\bibfnamefont {H.-H.}\ \bibnamefont {Wen}},
  \ and\ \bibinfo {author} {\bibfnamefont {R.}~\bibnamefont {Prozorov}},\
  }\href {\doibase http://dx.doi.org/10.1103/PhysRevB.89.144514} {\bibfield
  {journal} {\bibinfo  {journal} {Phys. Rev. B}\ }\textbf {\bibinfo {volume}
  {89}},\ \bibinfo {pages} {144514} (\bibinfo {year} {2014})}\BibitemShut
  {NoStop}%
\bibitem [{\citenamefont {Tanatar}\ \emph {et~al.}(2010)\citenamefont
  {Tanatar}, \citenamefont {Ni}, \citenamefont {Budâ€™ko}, \citenamefont
  {Canfield},\ and\ \citenamefont {Prozorov}}]{Tanatar2010SST}%
  \BibitemOpen
  \bibfield  {author} {\bibinfo {author} {\bibfnamefont {M.~A.}\ \bibnamefont
  {Tanatar}}, \bibinfo {author} {\bibfnamefont {N.}~\bibnamefont {Ni}},
  \bibinfo {author} {\bibfnamefont {S.~L.}\ \bibnamefont {Budâ€™ko}},
  \bibinfo {author} {\bibfnamefont {P.~C.}\ \bibnamefont {Canfield}}, \ and\
  \bibinfo {author} {\bibfnamefont {R.}~\bibnamefont {Prozorov}},\ }\href@noop
  {} {\bibfield  {journal} {\bibinfo  {journal} {Supercond. Sci. Technol.}\
  }\textbf {\bibinfo {volume} {23}},\ \bibinfo {pages} {054002} (\bibinfo
  {year} {2010})}\BibitemShut {NoStop}%
\bibitem [{\citenamefont {Prozorov}\ and\ \citenamefont
  {Giannetta}(2006)}]{Prozorov2006SST}%
  \BibitemOpen
  \bibfield  {author} {\bibinfo {author} {\bibfnamefont {R.}~\bibnamefont
  {Prozorov}}\ and\ \bibinfo {author} {\bibfnamefont {R.~W.}\ \bibnamefont
  {Giannetta}},\ }\href@noop {} {\bibfield  {journal} {\bibinfo  {journal}
  {Supercond. Sci. Technol.}\ }\textbf {\bibinfo {volume} {19}},\ \bibinfo
  {pages} {R41} (\bibinfo {year} {2006})}\BibitemShut {NoStop}%
\bibitem [{\citenamefont {Prozorov}\ \emph {et~al.}(2000)\citenamefont
  {Prozorov}, \citenamefont {Giannetta}, \citenamefont {Carrington},\ and\
  \citenamefont {Araujo-Moreira}}]{Prozorov2000PRB}%
  \BibitemOpen
  \bibfield  {author} {\bibinfo {author} {\bibfnamefont {R.}~\bibnamefont
  {Prozorov}}, \bibinfo {author} {\bibfnamefont {R.~W.}\ \bibnamefont
  {Giannetta}}, \bibinfo {author} {\bibfnamefont {A.}~\bibnamefont
  {Carrington}}, \ and\ \bibinfo {author} {\bibfnamefont {F.~M.}\ \bibnamefont
  {Araujo-Moreira}},\ }\href {<Go to ISI>://WOS:000088037000030
  http://prb.aps.org/pdf/PRB/v62/i1/p115_1} {\bibfield  {journal} {\bibinfo
  {journal} {Phys. Rev. B}\ }\textbf {\bibinfo {volume} {62}},\ \bibinfo
  {pages} {115} (\bibinfo {year} {2000})}\BibitemShut {NoStop}%
\bibitem [{\citenamefont {Fernandes}\ \emph {et~al.}(2012)\citenamefont
  {Fernandes}, \citenamefont {Vavilov},\ and\ \citenamefont
  {Chubukov}}]{TcEnhancement2012}%
  \BibitemOpen
  \bibfield  {author} {\bibinfo {author} {\bibfnamefont {R.~M.}\ \bibnamefont
  {Fernandes}}, \bibinfo {author} {\bibfnamefont {M.~G.}\ \bibnamefont
  {Vavilov}}, \ and\ \bibinfo {author} {\bibfnamefont {A.~V.}\ \bibnamefont
  {Chubukov}},\ }\href {\doibase 10.1103/PhysRevB.85.140512} {\bibfield
  {journal} {\bibinfo  {journal} {Phys. Rev. B}\ }\textbf {\bibinfo {volume}
  {85}},\ \bibinfo {pages} {140512} (\bibinfo {year} {2012})}\BibitemShut
  {NoStop}%
\bibitem [{\citenamefont {Kogan}(2009)}]{kogan2009PRB-2}%
  \BibitemOpen
  \bibfield  {author} {\bibinfo {author} {\bibfnamefont {V.~G.}\ \bibnamefont
  {Kogan}},\ }\href {\doibase 10.1103/PhysRevB.80.214532} {\bibfield  {journal}
  {\bibinfo  {journal} {Phys. Rev. B}\ }\textbf {\bibinfo {volume} {80}},\
  \bibinfo {pages} {214532} (\bibinfo {year} {2009})}\BibitemShut {NoStop}%
\bibitem [{\citenamefont {Homes}\ \emph {et~al.}(2004)\citenamefont {Homes},
  \citenamefont {Dordevic}, \citenamefont {Strongin}, \citenamefont {Bonn},
  \citenamefont {Liang}, \citenamefont {Hardy}, \citenamefont {Komiya},
  \citenamefont {Ando}, \citenamefont {Yu}, \citenamefont {Kaneko},
  \citenamefont {Zhao}, \citenamefont {Greven}, \citenamefont {Basov},\ and\
  \citenamefont {Timusk}}]{Homes2004Nature}%
  \BibitemOpen
  \bibfield  {author} {\bibinfo {author} {\bibfnamefont {C.~C.}\ \bibnamefont
  {Homes}}, \bibinfo {author} {\bibfnamefont {S.~V.}\ \bibnamefont {Dordevic}},
  \bibinfo {author} {\bibfnamefont {M.}~\bibnamefont {Strongin}}, \bibinfo
  {author} {\bibfnamefont {D.~A.}\ \bibnamefont {Bonn}}, \bibinfo {author}
  {\bibfnamefont {R.}~\bibnamefont {Liang}}, \bibinfo {author} {\bibfnamefont
  {W.~N.}\ \bibnamefont {Hardy}}, \bibinfo {author} {\bibfnamefont
  {S.}~\bibnamefont {Komiya}}, \bibinfo {author} {\bibfnamefont
  {Y.}~\bibnamefont {Ando}}, \bibinfo {author} {\bibfnamefont {G.}~\bibnamefont
  {Yu}}, \bibinfo {author} {\bibfnamefont {N.}~\bibnamefont {Kaneko}}, \bibinfo
  {author} {\bibfnamefont {X.}~\bibnamefont {Zhao}}, \bibinfo {author}
  {\bibfnamefont {M.}~\bibnamefont {Greven}}, \bibinfo {author} {\bibfnamefont
  {D.~N.}\ \bibnamefont {Basov}}, \ and\ \bibinfo {author} {\bibfnamefont
  {T.}~\bibnamefont {Timusk}},\ }\href@noop {} {\bibfield  {journal} {\bibinfo
  {journal} {Nature}\ }\textbf {\bibinfo {volume} {430}},\ \bibinfo {pages}
  {539} (\bibinfo {year} {2004})}\BibitemShut {NoStop}%
\bibitem [{\citenamefont {Martin}\ \emph {et~al.}(2009)\citenamefont {Martin},
  \citenamefont {Gordon}, \citenamefont {Tanatar}, \citenamefont {Kim},
  \citenamefont {Ni}, \citenamefont {Bud'ko}, \citenamefont {Canfield},
  \citenamefont {Luo}, \citenamefont {Wen}, \citenamefont {Wang}, \citenamefont
  {Vorontsov}, \citenamefont {Kogan},\ and\ \citenamefont
  {Prozorov}}]{Martin2009PRB}%
  \BibitemOpen
  \bibfield  {author} {\bibinfo {author} {\bibfnamefont {C.}~\bibnamefont
  {Martin}}, \bibinfo {author} {\bibfnamefont {R.~T.}\ \bibnamefont {Gordon}},
  \bibinfo {author} {\bibfnamefont {M.~A.}\ \bibnamefont {Tanatar}}, \bibinfo
  {author} {\bibfnamefont {H.}~\bibnamefont {Kim}}, \bibinfo {author}
  {\bibfnamefont {N.}~\bibnamefont {Ni}}, \bibinfo {author} {\bibfnamefont
  {S.~L.}\ \bibnamefont {Bud'ko}}, \bibinfo {author} {\bibfnamefont {P.~C.}\
  \bibnamefont {Canfield}}, \bibinfo {author} {\bibfnamefont {H.}~\bibnamefont
  {Luo}}, \bibinfo {author} {\bibfnamefont {H.~H.}\ \bibnamefont {Wen}},
  \bibinfo {author} {\bibfnamefont {Z.}~\bibnamefont {Wang}}, \bibinfo {author}
  {\bibfnamefont {A.~B.}\ \bibnamefont {Vorontsov}}, \bibinfo {author}
  {\bibfnamefont {V.~G.}\ \bibnamefont {Kogan}}, \ and\ \bibinfo {author}
  {\bibfnamefont {R.}~\bibnamefont {Prozorov}},\ }\href@noop {} {\bibfield
  {journal} {\bibinfo  {journal} {Phys. Rev. B}\ }\textbf {\bibinfo {volume}
  {80}},\ \bibinfo {pages} {020501} (\bibinfo {year} {2009})}\BibitemShut
  {NoStop}%
\bibitem [{\citenamefont {Ren}\ \emph {et~al.}(2008)\citenamefont {Ren},
  \citenamefont {Wang}, \citenamefont {Luo}, \citenamefont {Yang},
  \citenamefont {Shan},\ and\ \citenamefont {Wen}}]{RenWen2008PRL}%
  \BibitemOpen
  \bibfield  {author} {\bibinfo {author} {\bibfnamefont {C.}~\bibnamefont
  {Ren}}, \bibinfo {author} {\bibfnamefont {Z.-S.}\ \bibnamefont {Wang}},
  \bibinfo {author} {\bibfnamefont {H.-Q.}\ \bibnamefont {Luo}}, \bibinfo
  {author} {\bibfnamefont {H.}~\bibnamefont {Yang}}, \bibinfo {author}
  {\bibfnamefont {L.}~\bibnamefont {Shan}}, \ and\ \bibinfo {author}
  {\bibfnamefont {H.-H.}\ \bibnamefont {Wen}},\ }\href@noop {} {\bibfield
  {journal} {\bibinfo  {journal} {Phys Rev Lett}\ }\textbf {\bibinfo {volume}
  {101}},\ \bibinfo {pages} {257006} (\bibinfo {year} {2008})}\BibitemShut
  {NoStop}%
\bibitem [{\citenamefont {Li}\ \emph {et~al.}(2008)\citenamefont {Li},
  \citenamefont {Hu}, \citenamefont {Dong}, \citenamefont {Li}, \citenamefont
  {Zheng}, \citenamefont {Chen}, \citenamefont {Luo},\ and\ \citenamefont
  {Wang}}]{Li2008PRL}%
  \BibitemOpen
  \bibfield  {author} {\bibinfo {author} {\bibfnamefont {G.}~\bibnamefont
  {Li}}, \bibinfo {author} {\bibfnamefont {W.~Z.}\ \bibnamefont {Hu}}, \bibinfo
  {author} {\bibfnamefont {J.}~\bibnamefont {Dong}}, \bibinfo {author}
  {\bibfnamefont {Z.}~\bibnamefont {Li}}, \bibinfo {author} {\bibfnamefont
  {P.}~\bibnamefont {Zheng}}, \bibinfo {author} {\bibfnamefont {G.~F.}\
  \bibnamefont {Chen}}, \bibinfo {author} {\bibfnamefont {J.~L.}\ \bibnamefont
  {Luo}}, \ and\ \bibinfo {author} {\bibfnamefont {N.~L.}\ \bibnamefont
  {Wang}},\ }\href@noop {} {\bibfield  {journal} {\bibinfo  {journal} {Phys Rev
  Lett}\ }\textbf {\bibinfo {volume} {101}},\ \bibinfo {pages} {107004}
  (\bibinfo {year} {2008})}\BibitemShut {NoStop}%
\bibitem [{\citenamefont {Evtushinsky}\ \emph {et~al.}(2009)\citenamefont
  {Evtushinsky}, \citenamefont {Inosov}, \citenamefont {Zabolotnyy},
  \citenamefont {Viazovska}, \citenamefont {Khasanov}, \citenamefont {Amato},
  \citenamefont {Klauss}, \citenamefont {Luetkens}, \citenamefont
  {Niedermayer}, \citenamefont {Sun}, \citenamefont {Hinkov}, \citenamefont
  {Lin}, \citenamefont {Varykhalov}, \citenamefont {Koitzsch}, \citenamefont
  {Knupfer}, \citenamefont {Buchner}, \citenamefont {Kordyuk},\ and\
  \citenamefont {Borisenko}}]{EvtushinskyBoriesenko2009NJP}%
  \BibitemOpen
  \bibfield  {author} {\bibinfo {author} {\bibfnamefont {D.~V.}\ \bibnamefont
  {Evtushinsky}}, \bibinfo {author} {\bibfnamefont {D.~S.}\ \bibnamefont
  {Inosov}}, \bibinfo {author} {\bibfnamefont {V.~B.}\ \bibnamefont
  {Zabolotnyy}}, \bibinfo {author} {\bibfnamefont {M.~S.}\ \bibnamefont
  {Viazovska}}, \bibinfo {author} {\bibfnamefont {R.}~\bibnamefont {Khasanov}},
  \bibinfo {author} {\bibfnamefont {A.}~\bibnamefont {Amato}}, \bibinfo
  {author} {\bibfnamefont {H.-H.}\ \bibnamefont {Klauss}}, \bibinfo {author}
  {\bibfnamefont {H.}~\bibnamefont {Luetkens}}, \bibinfo {author}
  {\bibfnamefont {C.}~\bibnamefont {Niedermayer}}, \bibinfo {author}
  {\bibfnamefont {G.~L.}\ \bibnamefont {Sun}}, \bibinfo {author} {\bibfnamefont
  {V.}~\bibnamefont {Hinkov}}, \bibinfo {author} {\bibfnamefont {C.~T.}\
  \bibnamefont {Lin}}, \bibinfo {author} {\bibfnamefont {A.}~\bibnamefont
  {Varykhalov}}, \bibinfo {author} {\bibfnamefont {A.}~\bibnamefont
  {Koitzsch}}, \bibinfo {author} {\bibfnamefont {M.}~\bibnamefont {Knupfer}},
  \bibinfo {author} {\bibfnamefont {B.}~\bibnamefont {Buchner}}, \bibinfo
  {author} {\bibfnamefont {A.~A.}\ \bibnamefont {Kordyuk}}, \ and\ \bibinfo
  {author} {\bibfnamefont {S.~V.}\ \bibnamefont {Borisenko}},\ }\href@noop {}
  {\bibfield  {journal} {\bibinfo  {journal} {New J. Phys.}\ }\textbf {\bibinfo
  {volume} {11}},\ \bibinfo {pages} {055069} (\bibinfo {year}
  {2009})}\BibitemShut {NoStop}%
\end{thebibliography}

%

\end{document}